\documentclass[3p,numbers]{elsarticle}
\usepackage{amsmath}
\usepackage{amssymb}
\usepackage{nomencl}
\makenomenclature
 
\usepackage{graphicx}%
\usepackage{bm}%
\usepackage[usenames]{xcolor}
\usepackage{hyperref}%
\usepackage{subcaption} 
\usepackage{multicol} 
\usepackage{mathtools}

\newcommand{\qpr}{q_\text{pr}}
\newcommand{\xipr}{\xi_\text{pr}}
\newcommand{\rcyl}{\mathsf{r}}
\newcommand{\xcyl}{x}
\newcommand{\xapp}{\mathsf{x}}
\newcommand{\yapp}{\mathsf{y}}

\nomenclature{$\zeta_\text{a}(t)$}{quaternion-valued analytic signal of the problem. Introduced in \eqref{ansatz}, see also \cite{Ghirardo2018PRF}}
\nomenclature{$\Omega$}{spatial domain of the problem}
\nomenclature{$\chi(t)$}{nature angle, defined in \eqref{q}. The angle $2\chi$ is the latitude angle in Fig.~\ref{FigSphere}.a}
\nomenclature{$\omega_0$}{frequency of oscillation, at which the system would oscillate at constant amplitude and at zero growth rate, appearing in \eqref{dsdqw3a}}
\nomenclature{$\omega$}{thermoacoustic frequency, accounting for the effect of acoustic damping, flame response and nonlinear dynamics. See section \S\ref{secfst}}
\nomenclature{$\rho_0(\xcyl)$}{mean density, appearing in \eqref{fe2f2e2}}
\nomenclature{$\alpha$}{acoustic damping coefficient, appearing in \eqref{dsdqw}, defined in \eqref{dasdas13r3}, modelling non-reflective boundary conditions, volumetric damping and other losses}
\nomenclature{$\phi(t)$}{fast varying temporal phase, defined in \eqref{eqfastphi}} %
\nomenclature{$\varphi(t)$}{slowly varying temporal phase, defined in \eqref{q} and reintroduced in \eqref{eqc2}}%
\nomenclature{$\gamma$}{adiabatic index or heat capacity ratio, appearing in \eqref{fe2f2e2}}
\nomenclature{$\sigma$}{background noise intensity, appearing in \eqref{fe2f2e2}}
\nomenclature{$\theta$}{azimuthal coordinate in a cylindrical frame of reference, between $0$ and $2\pi$}
\nomenclature{$\theta_0(t)$}{see $n\theta_0$}
\nomenclature{$\xi(\theta,t)$}{stochastic source term $q_s$, projected on the axial mode of interest, appearing in \eqref{dsdqw1}, modelling the effect of non-coherent combustion noise. Defined in \eqref{aint4}, see also \S\ref{secstoc}}
\nomenclature{$\xipr(t)$}{stochastic source term $\xi$ projected on the mode shape $e^{in\theta}$. Defined in \eqref{ert1}, appearing in \eqref{midwife3}, complex valued}
\nomenclature{$\psi(\xcyl)$}{axial acoustic mode shape at the limit cycle, such that $\psi(\xcyl_b)=1$}
\nomenclature{$\mu_z(t)$}{quaternion valued noise term discussed in \eqref{stew}}
\nomenclature{$\nu$}{linear growth rate of the example presented in \S\ref{sec_isi}}
\nomenclature{$\kappa$}{nonlinear saturation constant of the describing function $Q$ considered in \S\ref{sec_isi}}

\nomenclature{$\varepsilon$}{formal smallness parameter in the fast stochastic differential equation \eqref{eq_2e1ww}}
\nomenclature[A2]{$A_p(\theta,t)$}{amplitude of pressure oscillation at the azimuthal angle $\theta$ at the flame position, introduced in \eqref{ampl_local2}. See also Fig.2.b in \cite{Ghirardo16JFM}}
\nomenclature[A1]{$A(t)$}{amplitude of oscillation, defined in \eqref{q}, non-negative. Radius of the sphere in Fig.~\ref{FigSphere}.a}
\nomenclature[B]{$B$}{integer number of noise terms in the averaged system \eqref{okok}. $B$ is set to $4$ after \eqref{eqDD22}}
\nomenclature[D1]{$D$}{in the main text, mean diameter of the annular combustor, appearing just before \eqref{dsdqw}. In \S\ref{savg1}, diffusion matrix of the Fokker-Planck equation of the averaged system, defined in \eqref{g0b}}
\nomenclature[D2]{$D_{\text{eq}}$}{equivalent diameter, such that a short, thin annular combustor with diameter $D_{\text{eq}}$ with fully reflective axial boundary conditions would have the same acoustic eigenfrequency of the studied annular combustor with partially reflective boundary conditions. See discussion after \eqref{fin2}}
\nomenclature[q1]{$q_1({\xcyl,\theta,t})$}{deterministic fluctuating heat release rate of a point in the whole domain $\Omega$, appearing in \eqref{fe2f2e2}}
\nomenclature[q2]{$q(\theta,t)$}{deterministic fluctuating heat release rate $q_1({\xcyl,\theta,t})$, projected on the axial mode of interest, appearing in \eqref{dsdqw1}, defined in \eqref{Qflame}}
\nomenclature[q3]{$\tilde{q}(\theta,t)$}{deterministic rescaled source term in the fluctuating pressure equation, accounting for both $q$ and the acoustic losses, defined in \eqref{def_11t}}
\nomenclature[q4]{$\qpr$}{deterministic source term $\tilde q$ in the fluctuating pressure equation, projected on the mode shape $e^{in\theta}$. Introduced in \eqref{ert0}, complex valued}
\nomenclature[q5]{${q_s}(\xcyl,\theta,t)$}{stochastic fluctuating heat release rate of a point in the spatial domain $\Omega$, modelling the effect of non-coherent combustion noise, appearing in \eqref{fe2f2e2}}
\nomenclature[q6]{$\mathcal{Q}_\theta[\,]$}{time-domain nonlinear operator modelling the source term $\tilde q$ appearing in \eqref{dsdqw3} as function of the azimuthal position $\theta$ in the annulus and of the local acoustic field. Introduced in \eqref{eqq1p}}
\nomenclature[q7]{$Q_\theta(A)$}{describing function of the nonlinear operator $\mathcal{Q}_\theta$, depending on the azimuthal position $\theta$ and on the local acoustic pressure amplitude $A$. Defined in \eqref{eqq1pQ}}
\nomenclature[T]{$T^\text{av}$}{time averaging operator, defined in \eqref{eqinreefe}}

\nomenclature[b]{$b$}{integer index with possible values $0,\ldots,B-1$}
\nomenclature[c1]{$c(\xcyl)$}{speed of sound, appearing in \eqref{fe2f2e2}}
\nomenclature[c1]{$\overline{c^2}$}{mean square of the speed of sound weighted on the mode shape, appearing in \eqref{dsdqw}, defined in \eqref{aint2}}
\nomenclature[c3]{$c_f$}{notation for $\cos(f)$, employed only in \S\ref{savg1}}
\nomenclature[f]{$f$}{drift function in the equations before averaging, appearing in its three terms in \eqref{rhseqeq} and in \eqref{eq_2e1ww}}
\nomenclature[g]{$g$}{diffusion function in the equations before averaging, appearing in \eqref{rhseqeq} and in \eqref{eq_2e1ww}}
\nomenclature[i]{$i,j,k$}{three imaginary units of the quaternion numbers, introduced after \eqref{ansatz}}
\nomenclature[m1]{$m$}{drift function in the averaged equations, appearing in \eqref{okok}, defined in \eqref{g0a}, and split into two terms in \eqref{ben4290342a0}}
\nomenclature[l]{$L$}{upper bound of the axial domain. See also $x$}
\nomenclature[m2]{$m_{v}^{\text{from}\,f}$}{part of the drift function $m$ arising from deterministic terms, which is present also in absence of noise, i.e.~ for $\sigma=0$. Defined in \eqref{eqtmp00a} and calculated in \S\ref{calc1}}
\nomenclature[m3]{$m_{v}^{\text{from}\,g}$}{part of the drift function $m$ arising from stochastic terms, which is present only for $\sigma\neq0$. Defined in \eqref{eqtmp00b} and calculated in \S\ref{calc2}}
\nomenclature[h]{$h$}{diffusion function in the averaged equations, appearing in \eqref{okok}, defined in \eqref{g0b2}, calculated in \S\ref{calc3}.}
\nomenclature[n]{$n$}{azimuthal order of the instability, positive integer, see \eqref{dsdqw} and \eqref{ansatz}} 
\nomenclature[n]{$n\theta_0(t)$}{azimuthal location of the pressure antinode of the standing component of the acoustic pressure. Introduced in \eqref{q} and then reintroduced in \eqref{eqc2}. It is the longitude angle of the system state on the Poincar\'e sphere of Fig.~\ref{FigSphere}.a, between $0$ and $2\pi$}
\nomenclature[p]{$p(\theta,t)$}{acoustic pressure at the axial location $\xcyl_b$ of the burners' exit in the combustion chamber. See \eqref{q}, \eqref{adep} and \eqref{ansatz}}
\nomenclature[p2]{$p_1(\xcyl,\theta,t)$}{acoustic pressure of a point in the spatial domain $\Omega$. See \eqref{fe2f2e2}, \eqref{adep}}
\nomenclature[r1]{$\rcyl$}{radial coordinate in a cylindrical frame of reference}
\nomenclature[r2]{$r$}{integer index with possible values $0,1$}
\nomenclature[s]{$s_f$}{notation for $\sin(f)$, employed only in \S\ref{savg1}}
\nomenclature[t]{$t$}{time variable}
\nomenclature[u1]{$u(\theta,t)$}{acoustic velocity in the azimuthal direction at the axial location $\xcyl_b$ of the burners' exit in the combustion chamber}
\nomenclature[u2]{$\tilde u(\theta,t)$}{rescaled acoustic velocity in the azimuthal direction, defined in \eqref{def_u1t}, see also \eqref{ee43432}, \eqref{eqc1}}
\nomenclature[u3]{${\bm u}_1(\xcyl,\theta,t)$}{acoustic velocity of a point in the domain $\Omega$. See \eqref{fe2f2e2}}
\nomenclature[vw]{$v,w$}{integer indices with possible values $0,1,2,3$}
\nomenclature[x1]{$\xcyl$}{axial coordinate in a cylindrical frame of reference, appearing in \eqref{adep}, between $0$ and $L$}
\nomenclature[x2]{$\xcyl_b$}{axial location just downstream of the burners, $0<\xcyl_b<L$} 
\nomenclature[x3]{$\xapp(t)$}{Displacement-equivalent accessory variable in \S\ref{savg1}, introduced in \eqref{xy}}
\nomenclature[y]{$\yapp(t)$}{velocity-equivalent accessory variable in \S\ref{savg1}, introduced in \eqref{xy}}
\nomenclature[Y1]{$\tilde Y$}{nondimensional acoustic admittance at the upstream and downstream boundary of the combustor domain, introduced in \eqref{bcadj}}
\nomenclature[Y2]{$Y(x)$}{scalar introduced in \eqref{DefY} defined only at $x=0$ and $x=L$. It matches $Y$ up to a sign difference}
\nomenclature[z1]{$z$}{generic quaternion-valued number}
\nomenclature[z2]{$z_v$}{generic $v$-th real-valued variable, introduced in \eqref{eq_2e1ww}}
\nomenclature{$\Lambda$}{norm of the axial mode shape $\psi(x)$. Introduced in \eqref{defLambda}}

\title{Averaging of thermoacoustic azimuthal instabilities}

\author{G. Ghirardo\corref{cor1}} 
\ead{giulio.ghirardo@ansaldoenergia.com}
\author{F. Gant}%
\address{Ansaldo Energia Switzerland, Haselstrasse 18, Baden 5400, CH}

\begin{document}

\date{\today}%

\begin{abstract}
We consider the acoustic flow field of rotationally symmetric systems, like an annular combustor and the flow in a round duct, in absence of a mean azimuthal flow field. We focus on azimuthal instabilities, which manifest as either spinning (rotating) waves or standing waves, or a linear combination of the two. These instabilities are often excited by some level of background noise that makes the system randomly change between spinning and standing states, undergo amplitude variations and changes to the azimuthal orientation of the solution. To account for this random change, we make use of a novel ansatz to track as a function of time the amplitude, orientation, nature (standing/spinning) and temporal phase of these instabilities. To capture the effect of the background noise, we apply stochastic averaging on the governing equations and obtain a novel differential equation. The equation allows to study the effects of acoustic sources and sinks on the statistics of the solution, and of explicit and spontaneous symmetry breaking and noise intensity. We focus on this last effect and show how the noise intensity affects the system preference for spinning and/or standing states. We find for example that, when present, background noise pushes the system away from spinning states and towards standing states, consistently with experiments and numerical simulations.
\end{abstract}

\maketitle
\newpage
\printnomenclature[1.8cm]

\section{Introduction}
\label{secintro}
In fluid dynamics, azimuthal instabilities occur as fluctuations of the pressure and velocity fields in geometries that exhibit a full rotational symmetry. This symmetry can be exact or approximate, as is common in real-world applications. We focus in particular on applications where the fluctuating field is acoustic in nature, but a connection with hydrodynamic instabilities is drawn in the conclusions. In this section, we briefly review the applications in \S\ref{secI1}, discuss the existing literature in \S\ref{Seclr}, and present the novelty of this work and the structure of the paper in \S\ref{secpa}.

\subsection{Typical applications}
\label{secI1}

Annular combustors exhibit a discrete rotational symmetry, \cite{Krueger99,Stow2001,Evesque2002,
Schuermans2010,
Bothien2015}, and can sustain thermoacoustic instabilities due to the coupling of the acoustics with the heat release rate of the flames \cite{Keller1995a}. These instabilities are often azimuthal of order $\pm n$, and travel at the speed of sound in the azimuthal direction in the annulus. %

Can-annular combustors consist instead of a set of equal cans, acoustically weakly communicating at the turbine inlet. In these systems azimuthal thermoacoustic instabilities can either occur as a set of synchronized states of pulsation between the different cans \cite{Bethke2002,Ghirardo2018JEGP,Ghirardo19JGTP}, or be localized in each round can separately, as perturbations of the acoustic modes of the round duct, at a frequency $\omega_0$ above the cut-on frequency of the duct. In this latter case, they are often azimuthal of order $n$ or mixed radial-azimuthal. %

In both applications, the system exhibits a pair of azimuthal eigenmodes that are degenerate, sharing the same eigenfrequency $\omega$, or close to degenerate, with very close frequencies. These modes are either linearly unstable or stable and close to the boundary of stability and are often subject to a random noise forcing term, due to turbulent heat release rate fluctuations.

We will neglect to consider situations where more than one thermoacoustic mode is excited at the same time, see e.g. \cite[Fig.~8]{Moeck2010b}. We will instead focus on the case where only one dominant peak at the frequency $\omega$ is present in the spectrum of the fluctuating field, plus other peaks at the multiples of $\omega$. We will also consider only the case where the mean velocity in the azimuthal direction is zero, i.e.~in absence of mean azimuthal flow, because in the applications mentioned in this introduction this is usually negligible (see e.g. \cite{Bauerheim2014b,Rouwenhorst2016} for  theoretical works discussing this effect). %

\subsection{Literature review}
\label{Seclr}
Because in this paper the focus is on capturing the instabilities as function of time, works that tackle the problem in the frequency domain only, e.g.~the works of \cite{Parmentier2012, Campa2014a, Mensah2016b} do not strictly apply.

In the time domain, one often chooses as ansatz for the acoustic field a linear combination of eigenmodes of the problem, where the coefficients of the combination depend on time and the eigenmode depends only on space. One often chooses as ansatz the combination of two spinning (rotating) waves travelling in the azimuthal direction, one clockwise and one counter-clockwise. In the same way, one can also express an ansatz as the combination of two standing waves with pressure antinodes fixed in space. Depending on the coefficients of these equivalent linear combinations, the solution describes a spinning state, a standing state, or a mixed state between the two. The choice of either basis depends on the application at hand. For the symmetric case, this is usually just a matter of preference, and many works \cite{Schuerm2006, Stow2009_foo, Ghirardo2013, Ghirardo16JFM} make use of standing modes as basis. Often systems that undergo an explicit symmetry breaking that depends on space are studied with a projection on standing modes \cite{Mehta2007,Noiray2012, Ghirardo2017JFM}, while systems with a mean azimuthal flow are studied with a projection on spinning waves \citep{Rouwenhorst2016, Hummel2016}, although there is no general rule on this, see e.g. \cite{Hagen2004,Cohen11}.

Annular combustors can be noisy environments, and experimental works with different level of noise are discussed later in \S\ref{sexper}. The effect of the noise can be observed in terms of how broad the probability density functions of the observed quantities are, see e.g.~Fig. 10 of \cite{Wolf2010}. The level of background noise $\sigma$ affects the dynamics of the fluctuating acoustic fields, which can then switch between standing and spinning states \cite{Krebs2002, Poinsot2011, Worth2013modaldyn}, as also presented in Fig.~\ref{FigSphere}. 

In some of the theoretical works earlier mentioned, the method of stochastic averaging is then applied to capture the effect of the noise. The resulting equations describe the slow time evolution of a three-dimensional phase space. The three dimensions are either $(A, B, \tilde\varphi)$ in the case of a projection on standing modes \cite{Schuerm2006} or $(F, G, \hat\varphi)$ in the case of a projection on spinning modes \cite{Rouwenhorst2016, Hummel2016}, where in both cases the first two variables are the amplitudes of the two modes and the third is their phase difference. However both phase spaces are ill-posed, i.e. there exist physical states to which more than one point correspond to in the phase space \citep{Ghirardo2018PRF}. Also, it is not possible to judge whether the system switches between standing or spinning states by just looking at the three variables $(A, B, \tilde\varphi)$, as was argued by \cite{Noiray2012}. To judge this, it is required instead to quantify how much the system is standing or spinning.

 \subsection{Proposed approach}
 \label{secpa}
\begin{figure}
  \centering
  \begin{subfigure}[t]{.34\columnwidth}
    \centering\includegraphics[width=\textwidth]{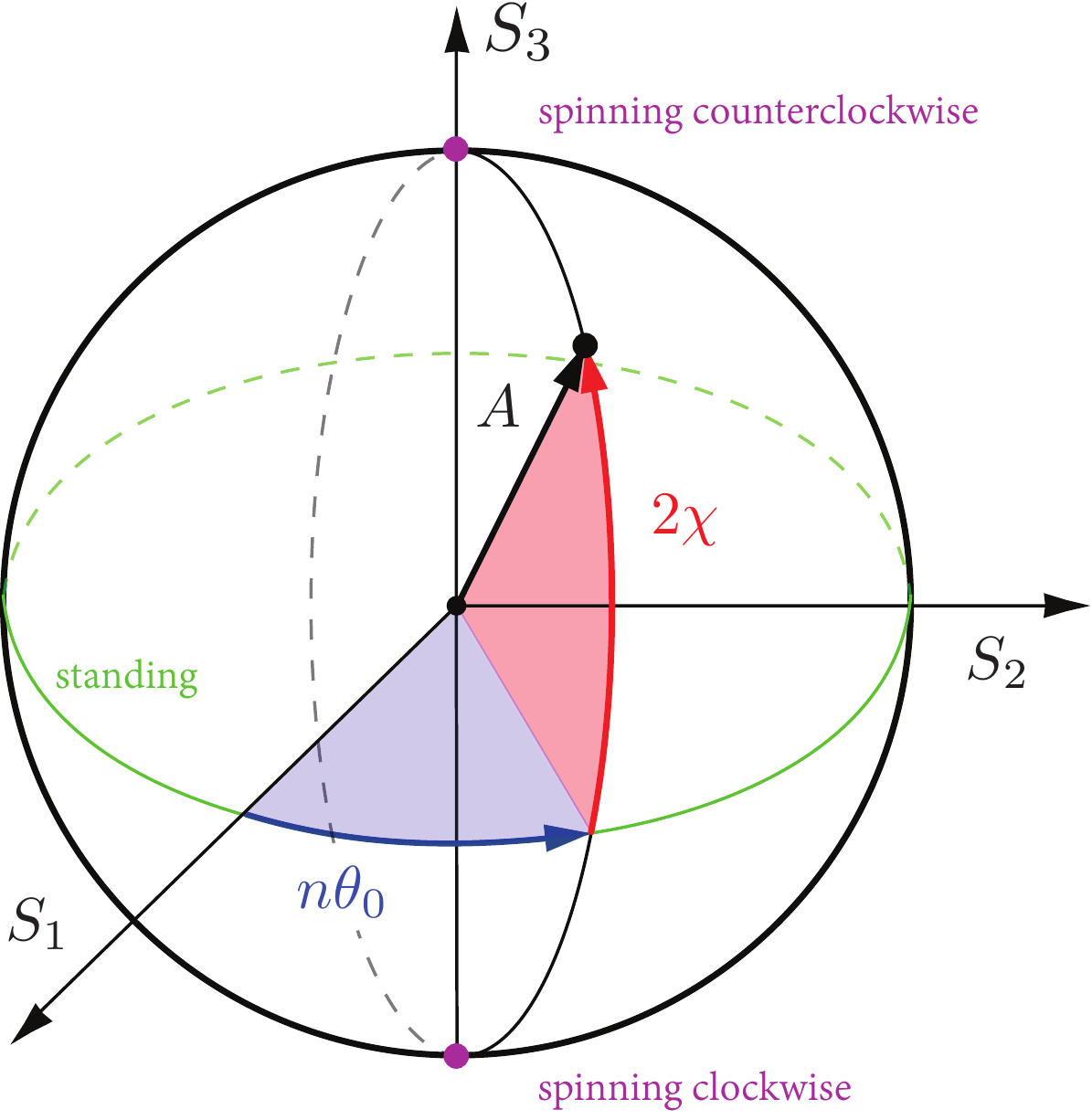}
    \caption{Poincar\'e representation}
  \end{subfigure}
  \begin{subfigure}[t]{.425\columnwidth}
    \centering\includegraphics[width=\textwidth]{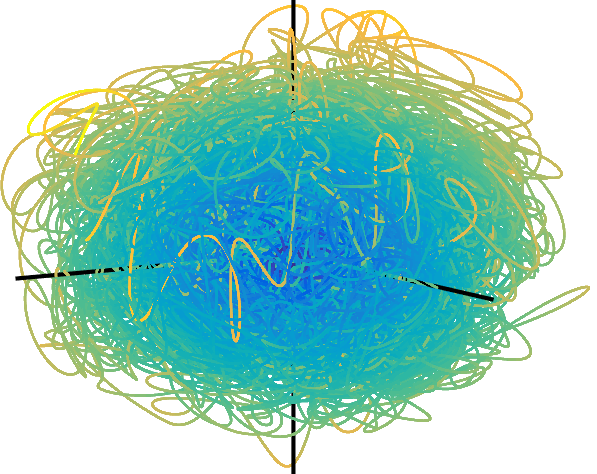}
    \caption{stochastic trajectories}
  \end{subfigure}
    \begin{subfigure}[t]{.185\columnwidth}
    \centering\includegraphics[width=\textwidth]{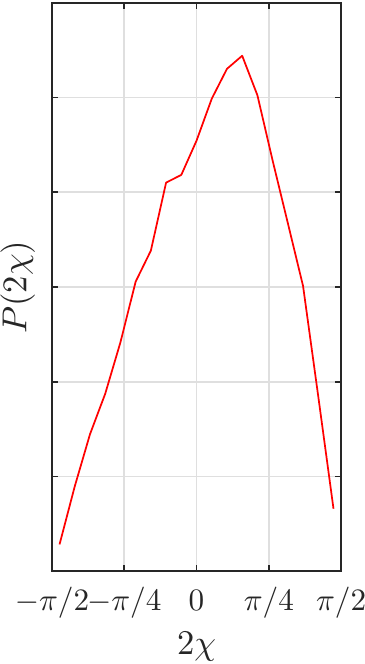}
    \caption{PDF of $2\chi$}
  \end{subfigure}
  \caption{a) Poincar\'e representation of the system state, characterized by the amplitude/radius $A$, the latitude angle $2\chi$ and the longitude angle $n\theta_0$. Points on the equator represent standing waves. Points at the north and south poles represent waves spinning respectively in the counterclockwise and clockwise direction. \label{FigSphere} In the same 3-dimensional space of a), we present in b) an example of the system state trajectory for the azimuthal instability of a nominally rotationally symmetric industrial annular combustor. The trajectory is described by the polar coordinates $(A(t),2\chi(t), n\theta_0(t))$ as function of time $t$ for approximately $130'000$ acoustic periods. The colour of the line describes qualitatively the amplitude $A$ of the point. c) probability density function (PDF) of the angle $2\chi$ for the trajectory in b), showing that the system is never in spinning states at $2\chi=\pm\pi/2$. This happens because of the background noise, which pushes the system away from the poles of the sphere in a), as discussed in this paper. Data were originally discussed in \cite{Ghirardo2018PRF}. }
\end{figure}

A quantitative indicator of the nature of the solution, i.e.~how much the system is standing or spinning, was devised by \cite{Bourgouin2013_asme} and called the spin ratio $s$. This indicator has been used only in experimental works so far, and only recently has been proven by \cite{Ghirardo2018PRF} that it is a state-space variable of the system, and not just an observable, as we discuss next. \cite{Ghirardo2018PRF} 
propose this new ansatz for the fluctuating field:
\begin{align}
\label{q}
p(\theta,t) =&A\cos(n(\theta-\theta_0))\cos(\chi)\cos(\omega t + \varphi) +A\sin(n(\theta-\theta_0))\sin(\chi)\sin(\omega t + \varphi)
\end{align}
In \eqref{q}, $p(\theta,t)$ is the fluctuating pressure field as function of the time $t$ and of the azimuthal coordinate $\theta$, the positive integer $n$ is the azimuthal order of the instability and the four variables $\{A,\theta_0,\chi,\varphi\}$ depend all on the time $t$. One advantage of \eqref{q} is that the variables $\{A,\theta_0,\chi,\varphi\}$ have a simple physical interpretation. $A$ is the non-negative amplitude of the solution. The orientation angle $n\theta_0$ is bounded in $[0, 2\pi)$ and is the location of the pressure antinode of the standing component of the acoustic field. The angle $\chi$ is bounded in $[-\pi/4,\pi/4]$ and is called the nature angle because it quantifies the nature of the solution, i.e.~whether the system is standing ($\chi=0$), spinning ($\chi=\pm\pi/4$), or in a state between standing and spinning. In particular, it is related to the spin ratio $s$ by the relation $s=\arctan(\chi)$. Finally, $\varphi$ is the slowly varying temporal phase of the oscillation. These four state-space variables $\{A,\chi,n\theta_0,\varphi\}$ can be used in low order models, as done later in \S\ref{stheory}, and can be reconstructed from numerical/experimental time series, allowing a direct validation of theoretical results. 
The three variables $\{A,n\theta_0,2\chi\}$ can be interpreted as spherical coordinates on the Poincar\'e sphere, as presented in Fig.~\ref{FigSphere}.a.%

To the knowledge of the authors, no theoretical nor experimental work to date has discussed the effect of the noise intensity $\sigma$ on the nature of the solution, i.e.~on whether the system prefers spinning or standing solutions. The ansatz \eqref{q}, where the nature angle $\chi$ is a state-space variable of the problem, offers this opportunity. It will be sufficient to characterize the dynamical behaviour of $\chi$ as a function of the noise intensity $\sigma$ to characterize the problem. We will show that noise pushes the system state away from spinning solutions.

We present in this paper a theory that allows studying azimuthal instabilities in the presence of noise and loss of rotational symmetry. In the simpler case where the system is rotationally symmetric and noise contribution is negligible, only the pure states (standing and spinning) can be limit-cycle solutions, and the theory reduces to existing theoretical criteria that prove, in agreement with the experiments, that either or both standing and spinning solutions can be stable limit-cycle, depending on certain conditions on the flame response \cite{Ghirardo16JFM, Laera2017a}. In the general case, a differential equation capturing the effect of noise and symmetry breaking on the amplitude $A$, orientation $n\theta_0$, phase $\varphi$ and nature angle $\chi$ of the azimuthal instabilities is discussed. This has already been published as a preprint by the authors in \cite{Ghirardo19a}, albeit without details.

In \S\ref{stheory} the ansatz \eqref{q} is substituted into the governing equations and a new differential equation is derived. The equation is discussed and general predictions on the effect of background noise are drawn. In \S\ref{sec_isi} the theory is applied to an academic problem, and with the help of numerical calculations the theoretical predictions are verified.
In \S\ref{sexper} a comparison with experimental evidence is presented, with good qualitative agreement. Conclusions are drawn in \S\ref{sconcl}.

\section{Theory}
\label{stheory}
In \S\ref{secgov} we briefly recall the fluctuating momentum and pressure equations for azimuthal instabilities. In \S\ref{sechrr1} and \S\ref{secstoc} we discuss the heat release rate model, based on the concept of nonlinear time-invariant operators. In \S\ref{sec1osc} we project the partial differential equations on the azimuthal mode of interest and obtain a complex-valued oscillator equation. In \S\ref{secavg22} we apply the method of stochastic averaging and discuss one by one the physical interpretation of the different terms in the resulting equations. In \S\ref{secchinoise} we focus on the effect of the noise and show that it pushes the system away from spinning states. In \S\ref{secff} we draw a parallel between the degenerate thermoacoustic instabilities discussed in this paper and degenerate hydrodynamic instabilities behind the turbulent wake of an axisymmetric object, showing how results discussed here apply there too.
\subsection{Governing equations}
\label{secgov}
The governing equations are the fluctuating pressure and momentum equations, assuming inviscid flow, linear acoustics, zero mean flow, as reviewed e.g.~by Clavin et al.~\cite[Appendix \S A]{Clavin94}:
\begin{subequations} 
\label{fe2f2e2}
\begin{align}
\label{eq_1}
\frac{\partial p_1}{\partial t} + \rho_0c^2\nabla\cdot{\bf{u}}_1&=(\gamma-1)q_1 +\sigma q_s\\
\label{eq_2}
\frac{\partial {\bf{u}}_1}{\partial t}+\frac{1}{\rho_0}\nabla p_1&=0
\end{align}
\end{subequations}
where a $0$ in the subscript denotes a steady quantity, e.g. $\rho_0$ is the mean density, and a $1$ in the subscript denotes unsteady quantities, e.g. $p_1$ is the acoustic pressure field and ${\bf{u}}_1$ is the acoustic velocity field. These two acoustic variables depend on the cylindrical coordinates $\xcyl,\rcyl,\theta$ and on the time $t$. To conclude the description of \eqref{fe2f2e2}, $c$ is the speed of sound and $\gamma$ is the adiabatic index. The right hand side of \eqref{eq_1} is the source term of the fluctuating pressure equation, which in thermoacoustics is the fluctuating heat release rate, in its deterministic component $q_1$ and its stochastic component $\sigma{q_s}$, both depending on $\{\xcyl,\rcyl,\theta,t\}$, respectively discussed later in \S\ref{sechrr1} and \S\ref{secstoc}.

We focus on systems that exhibit either a full rotationally symmetry, like a round duct, or a discrete rotational symmetry, like an annular combustor. We also consider systems that depart somewhat from this exact symmetry state, as is common in applications.

We consider the solution in cylindrical coordinates, and apply the method of separation of variables in the axial $\xcyl$, radial $\rcyl$ and azimuthal $\theta$ direction. We also assume for simplicity that the solution has a negligible dependence on the radial coordinate $\rcyl$, as the annulus is typically thin and azimuthal instabilities are to a very good approximation constant in the radial direction \cite{Munjal87}, even in laboratory annular combustors that exhibit a large annulus thickness compared to industrial combustors. We will also assume that the density $\rho_0$, and then also the speed of sound $c$, depend on the axial coordinate $x$ only. This allows to capture the steep gradient of temperature, density and velocity across the flame in the axial direction. We choose as ansatz for the acoustic pressure:
\begin{align}
\label{adep}
p_1(\xcyl,\theta,t) &= p(\theta,t)\psi(\xcyl)
\end{align}
where $\psi(\xcyl)$ describes the mode shape in the axial direction of the corresponding acoustic problem, and \eqref{adep} is interpreted as a Galerkin series expansion of the solution truncated to the first leading term \cite{Morse1953a,Culick2006,Ghirardo2017JFM}. Without any loss of generality, we set the value of the axial shape $\psi$ to one at the axial location $\xcyl_b$ just downstream of the burners in the combustion chamber, so that from \eqref{adep} $p(\theta,t)$ is the acoustic pressure field at the same axial location. 

The governing equations \eqref{fe2f2e2} are two-dimensional in space in $\xcyl$ and $\theta$. In \S\ref{apptrivial} it is shown how there exists a simpler set of equations that is one-dimensional in space in the azimuthal coordinate $\theta$, which is equivalent to the equations \eqref{fe2f2e2} once the ansatz \eqref{adep} is chosen. These simpler, equivalent equations can be obtained here informally by retaining only the azimuthal component of the momentum equation \eqref{eq_2}, for which $(\nabla\cdot {\bf{u}}_1)_\theta\approx2/D\partial u/\partial \theta $ where $u$ is the azimuthal component of the velocity $\bf{u}_1$:
\begin{subequations}
\label{dsdqw}
\begin{align}
\label{dsdqw1}
\frac{\partial p}{\partial t} + \frac{2\rho_0(\xcyl_b) \overline{c^2} }{D_\text{eq}}\frac{\partial u}{\partial \theta}&=(\gamma-1)q-\alpha p +\sigma\xi\\
\frac{\rho_0(\xcyl_b)D}{2n}\frac{\partial u}{\partial t}+\frac{1}{n}\frac{\partial  p}{\partial \theta}&=0
\end{align}
\end{subequations}
where we multiplied both sides of \eqref{eq_2} by $\rho_0D/2n$, and $n$ is the order of the azimuthal instability under consideration. The mode shape $\psi$ does not appear in \eqref{dsdqw}, because \eqref{dsdqw} characterize the acoustic field at the location $\xcyl_b$ just downstream of the burners, where the mode shape $\psi(\xcyl_b)=1$ simplify. To account for the axial extent of the acoustic field, in \eqref{dsdqw1} the equivalent diameter $D_\text{eq}$ substitutes the geometrical diameter $D$, and the mean square of the speed of sound $\overline{c^2}$ is found in place of $c^2$. Similarly on the right hand side of \eqref{dsdqw1} the additional term $-\alpha p$ appears because of the acoustic losses at the inlet and outlet of the combustor and/or other acoustic losses like volumetric damping. Completing the description of \eqref{dsdqw1}, $q$ denotes the projection of the heat release rate $q_1$ in \eqref{eq_1} on the azimuthal coordinate $\theta$ only, and similarly for $\xi$ in place of ${q_s}$. All equivalent and projected quantities are defined in the appendix in \S\ref{secspatavg}. We introduce for convenience the rescaled acoustic velocity $\tilde u$ and rescaled source term $\tilde q$ as
\begin{subequations}
\label{vybunim}
\begin{align}
\label{def_u1t} 
\tilde u &\coloneqq \frac{\rho_0(\xcyl_b) D}{2n} u\\
\label{def_11t}
\tilde q &\coloneqq (\gamma -1)q-\alpha p
\end{align}
\end{subequations}
where we use the symbol $\coloneqq$ to define the quantity on the left hand side. We substitute \eqref{vybunim} into \eqref{dsdqw} and obtain
\begin{subequations}
\label{dsdqw3}
\begin{align}
\label{dsdqw3a}
\frac{\partial p}{\partial t} + \omega_0^2\frac{1}{n}\frac{\partial \tilde u}{\partial \theta}&=\tilde q +\sigma\xi\\
\label{dsdqw3b}
\frac{\partial \tilde u}{\partial t}+\frac{1}{n}\frac{\partial  p}{\partial \theta}&=0
\end{align}
\end{subequations}
where $\omega_0\coloneqq 2n\sqrt{\overline{c^2}/(DD_\text{eq})}$ is the acoustic frequency of the azimuthal instability of interest when acoustic sources and sinks are set to zero.
For a thin annular chamber with wall-like boundary conditions and a mode shape $\psi(x)$ homogeneous in the axial direction, $D_\text{eq}$ matches the mean diameter of the annulus $D$.
Eq.~\eqref{dsdqw3} is a system of two coupled partial differential equations in the unknowns $p$ and $u$ as function of the independent variables time $t$ and azimuthal angle $\theta$. We discuss next individually the two source terms on the right hand side of \eqref{dsdqw3a}, to then come back to \eqref{dsdqw3} in \S\ref{sec1osc}.

\subsection{The deterministic flame response model}
\label{sechrr1}
We assume that the fluctuating heat release rate $q_1(\xcyl,\rcyl,\theta,t)$ appearing in \eqref{eq_1} depends on the position $\theta$ and on the acoustic field at the same position $\theta$. Conversely this then applies also to $q(\theta,t)$ appearing in \eqref{dsdqw1} and to $\tilde q(\theta,t)$ appearing in \eqref{dsdqw3a}: 
\begin{align}
\label{eqq}
\tilde q(\theta,t) = \mathcal{Q}_\theta\left[p(\theta, t),u(\theta, t)\right]
\end{align}
$\mathcal{Q}_\theta$ is a nonlinear, time-invariant operator that depends parametrically on the azimuthal coordinate $\theta$ to describe systems where burners positioned at different azimuthal locations have a different flame response, e.g.~\cite{Moeck2010b,Bauerheim2014JFM}. It also depends on the acoustic field just upstream of the flames, in terms of both acoustic pressure $p$ and azimuthal acoustic velocity $u$ as briefly discussed next.

The acoustic pressure $p$ in the combustion chamber excites, by means of the acoustic impedance of the whole system just upstream of the flame, the axial acoustic velocity $u_{\text{axial}}$ just upstream of the flame and in general the axial acoustic velocity in the burner. This velocity $u_{\text{axial}}$ is often used as a reference acoustic quantity by means of which the heat release rate $q_1$ is expressed \cite{Fleifil96,Schuermans99,Ducruix2000,Polifke2001a,Bellucci01,Lieuwen2003,Preetham2008a,Palies2010,Candel2013}. One can however exploit the link between $p$ and $u_{\text{axial}}$ and express the fluctuating heat release rate $q_1=q_1(u_{\text{axial}}(p))=q_1(p)$ as function of the acoustic pressure $p$ \cite{Ghirardo16JFM,Ghirardo2017b}, as verified experimentally by \cite{Gaudron2019}. The response of the flame to fluctuations of the acoustic pressure $p$ in the annular chamber is the main driver of the instability (see e.g. Fig. 18 in \cite{Wolf2012}), and are present in the linear and in the nonlinear regime.

The azimuthal acoustic velocity $u$ sweeps the flames transversally \cite{Connor2010,Hauser2011,OConnor2012,Connor2013a,Worth2020} and induces negligible axial velocity fluctuations \cite{Blimbaum2012}. Ghirardo and Juniper \cite{Ghirardo2013} consider a heat release model where the dependence on $u$ is nonlinear, and show that such dependence affects the nature angle $\chi$ and the stability of standing and spinning limit cycle solutions. It is proven theoretically by Acharya and Lieuwen \cite{Acharya2014} that, in absence of a mean azimuthal flow and for axisymmetric flames, the dependence on $u$ can in fact only be nonlinear. Consistently, experiments show that in the nonlinear regime both the amplitude and the phase of $u$ affect the flame response \cite{Saurabh2017,Saurabh2017a,Saurabh2019}. We refer to \cite{Ghirardo20ASME,Ghirardo20combsymp} for two recent works modelling this nonlinear effect. In the following we neglect this dependence on the azimuthal velocity and employ a simpler formulation where $\mathcal{Q}_\theta$ depends only on the acoustic pressure $p$:
\begin{align}
\label{eqq1p}
\tilde q(\theta,t) = \mathcal{Q}_\theta\left[p(\theta, t)\right]
\end{align}
A useful notion used in the following is the concept of describing function $Q_\theta$ of the operator $\mathcal{Q}_\theta$. For the single-input operator \eqref{eqq1p} this is defined as \citep{Gelb68}:
\begin{align}
\label{eqq1pQ}
Q_\theta(A,\omega)\coloneqq \frac{1}{A}\frac{1}{\pi/\omega}\int_0^{2\pi/\omega} \mathcal{Q}_\theta\left[A\cos(\omega t)\right] e^{-j\omega t}dt
\end{align}
In the definition \eqref{eqq1pQ}, an acoustic pressure sinusoid with amplitude $A$ and frequency $\omega$ is substituted as the argument of $\mathcal{Q}_\theta$. The function $Q_\theta(A,\omega)$ is called the sinusoidal-input describing function of the operator $\mathcal{Q}_\theta$, and quantifies the response of the operator $\mathcal{Q}_\theta$ at the frequency $\omega$ and input amplitude $A$. $Q_\theta(A,\omega)$ is an extension of the concept of the transfer function $Q_\theta(\omega)$, where the operator $\mathcal{Q}_\theta$ is not assumed to be linear in the amplitude $A$, and the nonlinear dependence on the amplitude $A$ is accounted for. We leave to a future investigation the discussion of the results in terms of the double-input model \eqref{eqq}, and employ in the following the model \eqref{eqq1p} that depends only on the acoustic pressure and its describing function \eqref{eqq1pQ}.

Since the focus of this paper is on the dynamics of one azimuthal instability at an approximately constant thermoacoustic frequency $\omega$, we fix $\omega$ in $Q_\theta$ and neglect to mention the dependence on $\omega$ in the following, as done elsewhere \citep{Schuerm2006,Noiray2011,Ghirardo2013}.
  If one focuses on one azimuthal mode with a wavelength much larger than the spacing between consecutive burners in the azimuthal direction, a distributed heat release model can be used, as carried out for example in \cite{Schuerm2006, Noiray2011, Ghirardo2013}, and later in \S\ref{sec_isi}. 
 If instead one considers the flames to be compact sources, the integral over $\theta$ becomes a summation over each flame, as done in \cite{Stow2009_foo, Ghirardo16JFM}.

These hypotheses on the deterministic flame response allow the derivation of analytical expressions for the deterministic part of the flame response but do not affect the results on the stochastic part of the flame response, which is discussed next.

\subsection{The stochastic flame contribution}
\label{secstoc}
In the last term on the right hand side of \eqref{eq_1}, ${q_s}(\xcyl,\theta,t)$ models the random component of the fluctuating heat release rate due to turbulent fluctuations at the flame, and the real valued $\sigma$ denotes its intensity \citep{Strahle1971,Strahle1972,Chiu1974,Rajaram2009}. We assume that this noise source term is additive and delta correlated in space and time, as modelled for example by \cite{Culick1992a}. The projection of ${q_s}(\xcyl,\theta,t)$ on the azimuthal direction leads to the source term $\xi(\theta, t)$ that appears in \eqref{dsdqw} and \eqref{dsdqw3}. We assume also that $q_s$ and $\xi$ do not depend on $\theta$. This allows to manipulate the projection of $\xi$ on the mode shape $e^{in(\theta-\theta_0)}$ as follows:
\begin{align}
\label{ert1}
{\xipr(t)} \coloneqq&\frac{1}{\pi}\int_0^{2\pi} e^{in(\theta-\theta_0)}\xi(\theta,t) d\theta= \xi_0(t)+i\xi_1(t)
\end{align}
The complex-valued quantity $\xipr$ is called the projected noise because it is the spatial projection of the noise field $\xi(\theta,t)$ on the azimuthal mode shape $e^{in(\theta-\theta_0)}$, and will appear later in the text. Because the field is random and we assume that $\xi$ does not depend on $\theta$, the projections on $\cos(n\theta)$ and on $\sin(n\theta)$ are two independent noise processes $\xi_0(t)$ and $\xi_1(t)$ that are white Gaussian sources with the same unit variance. Eq.~\eqref{ert1} holds regardless of the value of $n\theta_0$ because one can make the change of variable $\theta-\theta_0\rightarrow\tilde\theta$ in the integral and obtain the same result. In an annular combustor, \eqref{ert1} is equivalent to assuming that each flame emits the same intensity of background heat release rate due to turbulent fluctuations, and that a distributed flame model can be used with regards to this stochastic component.

\subsection{Oscillator equation}
\label{sec1osc}
In this section we simplify the governing equations \eqref{dsdqw3} to the equation of an oscillator, by projecting them on the mode of interest. To do so, we choose this ansatz for the acoustic variables $p,\tilde u$, which was proposed in \cite{Ghirardo2018PRF}:
\begin{subequations}
\label{ansatz}
\begin{align}
\label{ee4343}
2p(\theta,t) &= e^{-in\theta} \zeta_\text{a}'(t) + \mbox{q.c.} = 2\mbox{Re}\left[e^{-in\theta} \zeta_\text{a}'(t)\right]\\
\label{ee43432}
2\tilde u(\theta, t) &=ie^{-in\theta} \zeta_\text{a}(t) + \mbox{q.c.}
\end{align}
\end{subequations}
where a prime denotes a time derivative. In the ansatz, the dependence on the azimuthal direction $\theta$ is encoded in the complex exponential, while $\zeta_\text{a}$ and its time derivative $\zeta_\text{a}'$ depend on time only, are quaternion valued, and are kept generic for the time being. We show later in \S\ref{secavg22} that \eqref{ansatz} is equivalent to the ansatz \eqref{q} for a suitable choice of $\zeta_\text{a}$. An introduction to quaternion algebra can be found in the book of Doran and Lasenby \cite{Doran2003}. A generic quaternion number can be written as $\zeta_\text{a}(t)=\zeta_{\text{a},0}(t)+i\zeta_{\text{a},1}(t)+j\zeta_{\text{a},2}(t)+k\zeta_{\text{a},3}(t)$, where the $\zeta_{\text{a},v}(t),\,v=0,1,2,3$ are real valued numbers and $i,j,k$ are three imaginary units defined by $i^2=j^2=k^2=-1$ and $ij=k$, $jk=i$, $ki=j$. In \eqref{ansatz}, the expression $\text{q.c.}$ denotes the quaternion conjugate of the quantity to its left: given a quaternion number $z=z_0+iz_1+jz_2+kz_3$, its quaternion conjugate is $z^*=z_0-iz_1-jz_2-kz_3$. In quaternion algebra the product of two quaternions is not commutative, so that the order of the complex exponential and of the quaternion number in \eqref{ansatz} matters. One can deduce from \eqref{ansatz} that the imaginary unit $i$ is linked to a spatial (azimuthal) phase information. We will discuss the physical interpretation of the other imaginary units $j,k$ when the ansatz for $\zeta_\text{a}$ is introduced. Exhaustive information on \eqref{ansatz} is presented in \citep{Ghirardo2018PRF}.

Since the ansatz \eqref{ansatz} solves \eqref{dsdqw3b} by construction, we focus on \eqref{dsdqw3a} next. We substitute \eqref{ansatz} into \eqref{dsdqw3a} multiplied by two and obtain
\begin{align}
\label{midwife}
e^{-in\theta}\left[\zeta_\text{a}''+\omega_0^2\zeta_\text{a}\right] + \mbox{q.c.} = 2\tilde q + 2\sigma \xi
\end{align}
We multiply \eqref{midwife} on the left by $e^{in\theta}$ and obtain
\begin{align}
\label{midwife2}
\zeta_\text{a}''+\omega_0^2\zeta_\text{a} + e^{in\theta}\left(\zeta_\text{a}''^* + \omega_0^2\zeta_\text{a}^*\right) e^{in\theta} = 2e^{in\theta}\tilde q + 2\sigma e^{in\theta}\xi
\end{align} 
where the asterisk denotes quaternion conjugation.
The third term on the left hand side of  \eqref{midwife2} originates from the $\mbox{q.c.}$ in \eqref{midwife}, exploiting that for any two quaternion numbers $a,b$ the property $(ab)^* = b^*a^*$ holds. We average both sides of \eqref{midwife2} over $\theta$ in $[0,\,2\pi)$ to obtain
\begin{align}
\label{midwife3}
\zeta_\text{a}''+\omega_0^2\zeta_\text{a} + \frac{1}{2\pi}\int_0^{2\pi}e^{in\theta}\left(\zeta_\text{a}''^* + \omega_0^2\zeta_\text{a}^*\right)  e^{in\theta} d\theta = \qpr + \sigma\xipr
\end{align}
where $\xipr$ was introduced in \eqref{ert1} and the projected heat release rate $\qpr$ is defined as:
\begin{align}
\label{ert0}
{\qpr} \coloneqq& \frac{1}{\pi}\int_0^{2\pi}e^{in\theta}\tilde q\,d\theta
\end{align}
We simplify next the left hand side of \eqref{midwife3} and observe that for any quaternion $z=z_0+iz_1+jz_2+kz_3$ not dependent on $\theta$ we have that
\begin{align}
\label{ppq}
z+\frac{1}{2\pi}\int_0^{2\pi}e^{in\theta} z^*  e^{in\theta} d\theta = z -jz_2 -kz_3 = z_0+iz_1 = \frac{z -izi}{2}\qquad \forall n=1,2,\ldots
\end{align}
We substitute the identity \eqref{ppq} with $z=\zeta_\text{a}''+\omega_0^2\zeta_\text{a}$ into the left hand side of \eqref{midwife3}, and then multiply both sides by $2$:
\begin{align}
\label{midwife4}
\zeta_\text{a}''+\omega_0^2\zeta_\text{a} -i\left[\zeta_\text{a}''+\omega_0^2\zeta_\text{a}\right]i = 2 {\qpr} + 2\sigma{\xipr}
\end{align}
Equation \eqref{midwife4} describes a complex-valued oscillator with natural frequency $\omega_0$, where $2\pi/\omega_0$ is the acoustic period of the acoustic mode of interest, which is the fast time scale of the problem. For this reason, we call $\zeta_\text{a}$ a fast oscillating variable. Refer to \cite{Cveticanin1992} for a discussion of a similar, complex-valued oscillator equation. The state of the system at an instant $t$ consists of the two real-valued components $\zeta_{\text{a},0}(t)$ and $\zeta_{\text{a},1}(t)$ and their two time derivatives.
\subsection{Averaged equations}
\label{secavg22}
In this section we map the oscillator equation \eqref{midwife4} in terms of the quaternion-valued variable $\zeta_\text{a}(t)$ to a new equation in terms of a novel ansatz. We then apply the method of stochastic averaging and interpret the resulting equation.

We introduce this ansatz for $\zeta'_\text{a}$:
\begin{align}
\label{eqc2}
\zeta'_\text{a}(t) = A(t)e^{in\theta_0(t)}e^{-k\,\chi(t)}e^{j(\omega t + \varphi(t))}
\end{align}
The variables $\{A,n\theta_0,2\chi\}$ are the same variables discussed in \S\ref{secpa} and can be interpreted as spherical coordinates on the sphere of Fig.~\ref{FigSphere}.a. From here onwards, for ease of notation we drop the explicit dependence of the four variables $\{A,n\theta_0,\chi,\varphi\}$ on the time $t$. We notice that \eqref{eqc2} and \eqref{ee4343} are fully equivalent to the ansatz \eqref{q} introduced in \S\ref{secintro}. In fact if we substitute \eqref{eqc2} into \eqref{ee4343} divided by two we recover the ansatz \eqref{q}. Equation \eqref{eqc2} allows the discussion of the physical interpretation of the three imaginary units. As discussed after \eqref{ansatz}, also in \eqref{eqc2} the imaginary unit $i$ encodes spatial (azimuthal) information. The imaginary unit $j$ appearing in $e^{j(\omega t + \varphi(t))}$ encodes a temporal phase. The imaginary unit $k$ appears only in relation to the nature angle $\chi$ and we can say that it encodes nature information, as $\chi$ is the nature angle. The imaginary units $i$ and $k$ can also be interpreted in relation to the Poincar\'e sphere representation of Fig.\ref{FigSphere}, where they encode longitude and latitude information respectively. The ansatz \eqref{eqc2} offers the following advantages in comparison with a standard projection on standing and spinning modes \cite{Ghirardo2018PRF}: \textit{1)} invariance of the representation with respect to a change of the frame of reference\footnote{except of course for the orientation angle $n\theta_0$, which depends on the azimuthal angle of the frame of reference}; \textit{2)} well-posedness of the phase space: the trajectory of the system on the Poincar\'e sphere of Fig. \ref{FigSphere}.b is always a continuous path; \textit{3)} direct physical interpretation of the state space variables appearing in the ansatz, without any need to resort to additional quantities to characterize the system in a physically meaningful way\footnote{for example, in order to discuss whether the system is standing or spinning, one would often resort to the spin ratio $s$, which needed to be first calculated from the system state}.

The next step is to substitute \eqref{eqc2} into \eqref{midwife4} and to apply the method of stochastic averaging. The application of the method is detailed in \S\ref{savg1}, and only an overview of the assumptions and a discussion is presented in the main text here. The method simplifies a set of differential equations that depend directly on the time $t$ and exhibit a fast oscillation in time that is modulated at a slower timescale. In the case at hand, the fast oscillation consists of the term $e^{j\omega t}$ in \eqref{eqc2}, which oscillates at the fast timescale of the thermoacoustic period of oscillation $2\pi/\omega$. The slow modulation consists of the variations in time of the four variables $\{A,n\theta_0,\chi,\varphi\}$, which are called slow variables because they change with a larger timescale than the period. One then averages in time the original equations over the period $2\pi/\omega$. The resulting simplified equations do not depend directly on the time $t$ (while in the original equations the dependence on $t$ is direct because of the term $e^{j\omega t}$ in \eqref{eqc2}) and describe the time derivatives of the slow variables. The accuracy of the averaged equations depends on how weakly nonlinear is the system. This means that in the governing equations the energy sources (usually the flames), the energy sinks (usually the acoustic damping), and the stochastic term should not be too large on the right-hand side of \eqref{midwife4}. This is often the case in thermoacoustic applications, as reviewed by \cite{Ghirardo2017b} in terms of linear growth rates of the whole systems, with flame switched on and off.
Under this assumption, averaging the complex-valued, second order equation \eqref{midwife4} leads to the following quaternion-valued equation for the slow variables:
\begin{align}
\nonumber
(\ln A)'+(n\theta_0'+&\varphi'\sin(2\chi))i+\varphi'\cos(2\chi)j-\chi'k = \\
\nonumber
&+\frac{1}{2}\frac{1}{2\pi}\int_0^{2\pi} \left(e^{i2n(\theta-\theta_0)}e^{k\chi}+e^{-k\chi}\right)Q_\theta(A_p(\theta))d\theta\,e^{k\chi}\\
\label{bb}
&+\left(-\frac{\omega}{2}+\frac{\omega_0^2}{2\omega}\right)e^{-k\chi}je^{k\chi}+\frac{\sigma^2}{4A^2}\left(1+\tan(2\chi)k\right)+\frac{\sigma}{\sqrt{2}A}\mu_z
\end{align}
Equation \eqref{bb} is discussed in the rest of this section, from left to right, for an arbitrary function $Q_\theta$. It is later simplified for an academic test case in \S\ref{sec_isi}.

\subsubsection{Left hand side}
The left hand side of \eqref{bb} characterizes the rate of change of the variables $\{A,n\theta_0,\varphi,\chi\}$, which all depend on the time $t$ only. The real part of this rate of change is $(\ln A)'=A'(t)/A(t)$, which describes how the amplitude changes with time. The $k$-imaginary part of the left-hand side of \eqref{bb} describes the rate of change of the nature angle $\chi$. The $i$-imaginary and $j$-imaginary parts of the left hand side of \eqref{bb} are not pure time derivative of quantities, because of the $\sin(2\chi)$ and $\cos(2\chi)$ terms. These two terms can, however, be interpreted parametrically in the nature angle $\chi$. For a fixed angle $\chi$, they describe together both the change of the location $n\theta_0$ of the pressure antinode of the standing component of the pressure field and the temporal phase $\varphi$. The variable $\chi$ directly affects to what extent the two derivatives $n\theta_0'$ and $\varphi'$ are related, because of the $\sin(2\chi)$ and $\cos(2\chi)$ terms.

If $\chi=0$, the $i$- and $j$- parts of the left hand side of \eqref{bb} describe the change of $n\theta_0$ and $\varphi$ respectively. In this case, by substituting $\chi=0$ into \eqref{q} we obtain:
\begin{align}
\label{dewnqd}
p(\theta,t) =&A\cos(n(\theta-\theta_0))\cos(\omega t + \varphi)
\end{align}
which corresponds to a standing mode. In the solution \eqref{dewnqd} of the standing mode, $n\theta_0$ is the spatial phase and $\varphi$ is a temporal phase.

If instead $\chi=\pm\pi/4$, substituting it into \eqref{q} we obtain
\begin{align}
p(\theta,t)
\label{dwqdqw}
=&\frac{A}{\sqrt{2}}\cos\left[\omega t\mp(n\theta-(n\theta_0\pm\varphi))\right]
\end{align}
In \eqref{dwqdqw} we observe that for a purely spinning state, the spatial phase $n\theta_0$ and the temporal phase $\varphi$ appear together in the same trigonometric function, with total  spatio-temporal phase $n\theta_0\pm\varphi$. The time derivative of this total phase is described by the $i$-imaginary part of the left hand side of \eqref{bb} for this spinning case.

In the general case in which the system is not in a pure spinning or standing state, the rates of change $n\theta_0'$ and $\varphi'$ are linked by the nature angle $\chi$.
\subsubsection{Fluctuating heat release rate and acoustic losses term}
\label{sechrr2}
The first term on the right hand side of \eqref{bb} describes the flame response:
\begin{align}
\label{termQ}
\frac{1}{2}\frac{1}{2\pi}\int_0^{2\pi} \left(e^{i2n(\theta-\theta_0)}e^{k\chi}+e^{-k\chi}\right)Q_\theta(A_p(\theta))d\theta\,e^{k\chi}
\end{align}
where the describing function $Q_\theta$ was introduced in \eqref{eqq1pQ}, and depends nonlinearly on the local amplitude $A_p$ of the acoustic pressure at the flame location $\theta$:
\begin{align}
\label{ampl_local2}
A_p(\theta) &\coloneqq A\sqrt{\cos^2(n\theta-n\theta_0)\cos^2(\chi)+\sin^2(n\theta-n\theta_0)\sin^2(\chi)}
\end{align}
The amplitude $A_p$ was introduced (as function of different variables) in \cite[called $R$ in their eq. (3.6)]{Ghirardo16JFM} and exemplified for standing and spinning states in Fig. 2 therein. If the flame model \eqref{eqq} were used, the describing function $Q_\theta$ would depend more generally on all three variables $A,n\theta_0,\chi$. This would allow one for example to model the dependence of the heat release rate on the nature angle $\chi$, as suggested by \cite{Nygard2018}, who concluded that \textit{different flame responses may need to be defined in order to characterize CW and ACW spinning modes}. The describing function response is integrated over the annulus in the azimuthal variable $\theta$, to account for the contribution of all the flames.

\subsubsection{Frequency shift term}
\label{secfst}
The second term on the right hand side of \eqref{bb} is $\left(-\frac{\omega}{2}+\frac{\omega_0^2}{2\omega}\right)e^{-k\chi}je^{k\chi}$. This term has only $i$- and $j$-imaginary non-zero parts. This means that in the equations, together with the part of the heat release rate not in phase with the pressure, affect the rates of change of $\varphi$ and $n\theta_0$, which appear in the $i$- and $j$- imaginary parts. To determine the mean frequency of thermoacoustic oscillation $\omega$, one looks for a value $\omega$ such that the mean value of $\varphi'$ is zero, simply because a steady drift trend like $\varphi=\Delta\omega\, t$ is by definition an increase $\Delta\omega$ of frequency. One then sets to zero the derivative $\varphi'$ and looks for a value of $\omega$ solution of the governing equations, averaged over an observation time longer than the acoustic period.
When making predictions, usually the acoustic frequency $\omega_0$ is known and one estimates the thermoacoustic frequency $\omega$ as just described. When processing experimental time series, $\omega$ can be calculated as the mean slope of the reconstructed phase, while $\omega_0$ is not known.
\subsubsection{Deterministic term due to the noise}
\label{secss}
The third term on the right hand side of \eqref{bb} is $\frac{\sigma^2}{4A^2}\left(1+\tan(2\chi)k\right)$. It is easier to discuss the term in its two parts, real and $k$-imaginary.

The real part is a force acting on the variable $\ln(A)'$ on the left-hand side. %
This term is always positive and then pushes the system to higher amplitudes, and is largest close to the origin $A=0$, where it tends to $\infty$. It is the same term appearing for axial instabilities.

The $k$-imaginary part $\sigma^2\tan(2\chi)/4A^2$ becomes singular not just at $A=0$, but also at $\chi=\pm\pi/4$, and is one key result of the paper, discussed later in \S\ref{secchinoise}.
This is exemplified in the example of \S\ref{sec_isi}.

\subsubsection{Stochastic term due to the noise} 
\label{stew}The last term on the right hand side of \eqref{bb} is $\frac{\sigma}{\sqrt{2}A}\mu_z = \frac{\sigma}{\sqrt{2}A}\left(\mu_0+i\mu+j\mu_2+k\mu_3\right)$, where $\mu_z(t)$ is a quaternion-valued random stochastic process, and $\mu_v(t)$ for $v=0,1,2,3$ are independent, real-valued white Gaussian processes with unit variance. Their cross-correlation function is $R_{vw}(\tau)=\delta(\tau)\delta_{vw}\,\,\,\forall v,w \in \{0,1,2,3\}$ where $\tau$ is the correlation time, $\delta(\tau)$ is the Dirac distribution and $\delta_{vw}$ is the Kronecker delta. The whole term scales linearly with the background noise intensity $\sigma$.

\subsection{Effect of the background noise $\sigma$ on the nature angle}
\label{secchinoise}
The dynamics of the nature angle $\chi$ are described by the $k$-imaginary part of the system \eqref{bb}:
\begin{align}
\label{eqchichichi2}
\chi'&=\left[\frac{1}{4\pi}\int_0^{2\pi} \left(e^{i2n(\theta-\theta_0)}e^{k\chi}+e^{-k\chi}\right)Q_\theta(A_p(\theta))d\theta\,e^{k\chi}\right]_k-\frac{\sigma^2\tan(2\chi)}{4 A^2}+\frac{\sigma\mu_3}{\sqrt{2}A}
\end{align}
where the subscript on the square bracket denotes the $k$-imaginary part.

In the deterministic case, $\sigma$ is zero, and only the first term on the right hand side of \eqref{eqchichichi2} is left. This term then determines alone the fixed points of $\chi$, i.e.~whether standing ($\chi=0$) and spinning ($\chi=\pm\pi/4$) states are fixed points of \eqref{eqchichichi2} and if they are stable or not. It describes the effect of explicit symmetry breaking, where $Q$ directly depends on $\theta$, and spontaneous symmetry breaking, which occurs due to the term $e^{i2n(\theta-\theta_0)}$ and was discussed in detail by \cite{Ghirardo16JFM} leading to the $N_{2n}$ criterion for the stability of standing solutions.

In the stochastic case, i.e.~when the level $\sigma$ of background noise is not negligible, the last two terms in \eqref{eqchichichi2} are not zero. They are functions of the ratio $\sigma/A$ and discussed next separately. The second to last term $-\sigma^2\tan(2\chi)/4 A^2$ tends to $\mp\infty$ for $\chi\rightarrow\pm\pi/4$, and then spinning states $\chi=\pm\pi/4$ cannot be fixed points of the system in the stochastic case. This term pushes the system state towards positive values when $\chi$ is negative, and towards negative values when $\chi$ is positive. This effect is proportional to the square of the noise intensity $\sigma$. In other words, the noise pushes the state variable $\chi$ away from the boundaries $\pm\pi/4$ of the domain for $\chi$, which are spinning states, towards standing states at $\chi=0$, as in Fig.~\ref{FigSphere}.c.

The last term on the right hand side of \eqref{eqchichichi2} is large at  small amplitudes, at which the nature angle of the system can change very quickly. This is expected, since when the pressure field is really small even a small perturbation can completely change the solution, and hence the nature angle $\chi$. However, as the system state approaches the spinning states at $\chi=\pm\pi/4$, the second-to-last term dominates because it scales like $A^{-2}$. This term guarantees that the variable $\chi$ stays in the bounded domain $(-\pi/4,\pi/4)$.

\subsection{Applicability to hydrodynamic instabilities behind a turbulent axisymmetric wake}
\label{secff}
In this section, we draw a link with hydrodynamic azimuthal instabilities, in particular with the problem of a wake behind an axisymmetric round object. This problem exhibits oscillating azimuthal hydrodynamic instabilities of order $m=\pm1$ above a critical Reynolds number at which a Hopf bifurcation occurs \cite{Mittal2002, Fabre2008}. These global modes are large-scale vortex-shedding motions, which 
persist also at high Reynolds numbers \cite{Rigas2014a}, where turbulence acts as a noise source in the governing equations \cite{Rigas2015}. The transferability of the results presented in this paper to the case of these hydrodynamic instabilities should follow from normal form theory \cite{Crawford1991}, though the actual equations would need to be re-derived for the hydrodynamic case. The scaling of the equivalent noise intensity $\sigma$ as a function of the Reynolds number is deemed of interest, ultimately allowing to discuss the preference of hydrodynamic instabilities for standing and spinning states as a function of Reynolds.

\section{A numerical example}
\label{sec_isi}
In this section we consider an example to which the averaged equations \eqref{bb} apply. The example is of relevance in applications, and allows us both to verify numerically the predictions of the previous section, and to present a simple case for which the equations simplify.

We consider an annular combustor that is rotationally symmetric and a distributed heat release rate model, for increasing values of the noise intensity $\sigma$. We choose this expression for $\tilde q$ in \eqref{def_11t}:
\begin{align}
\label{modfuff}
\tilde q[p(t)]=2\nu p(t)-\kappa p^3(t)
\end{align}
In this case, the coefficient $\nu$ is the linear growth rate of the system and can be written as the difference of $\nu=\nu_{\text{hrr}}-\alpha/2$ where $\nu_{\text{hrr}}$ accounts for the positive contribution of the heat release rate as energy source. Equation \eqref{modfuff} is the simplest heat release rate model possible and can be interpreted as the truncation to the third order of a power expansion of any heat release model that has a structure like $q=\mathcal{Q}[p]$, see e.g. \cite{Lieuwen2003a,Eisenhower2008,Noiray2011} where a similar polynomial expansion has been applied to thermoacoustic problems. The model \eqref{modfuff} also accounts only for the part of the heat release rate that is in phase with the pressure $p$, which contributes to the Rayleigh criterion. We also restrict to the more common case of a supercritical instability, i.e.~assume that $\kappa>0$, neglecting to consider the subcritical case.

The describing function $Q$ of the time domain operator \eqref{modfuff} is calculated by means of \eqref{eqq1pQ}:
\begin{align}
\label{Qdf}
Q(A) = 2\nu - \frac{3}{4}\kappa A^2
\end{align}
where we dropped the subscript $\theta$ from $Q$ since we assume rotational symmetry. Substituting \eqref{ampl_local2} into \eqref{Qdf}, and in turn \eqref{Qdf} into \eqref{termQ}, after some manipulations we obtain:
\begin{align}
\label{termQ2}
\frac{1}{2}\frac{1}{2\pi}\int_0^{2\pi} \left(e^{i2n(\theta-\theta_0)}e^{k\chi}+e^{-k\chi}\right)Q(A_p(\theta))d\theta\,e^{k\chi}=\nu-\frac{3}{16}\kappa A^2-\frac{3}{32}\kappa A^2\cos(2\chi)e^{2\chi k}
\end{align}
We substitute \eqref{termQ2} into \eqref{bb}. The real and $k$-imaginary parts of the resulting equation are:
\begin{align}
\label{dwew11}
(\ln A)'-\chi' k = \nu-\frac{3\kappa}{16}A^2-\frac{3\kappa}{32}A^2\cos(2\chi)e^{2\chi k}+\frac{\sigma^2}{4A^2}\left(1+\tan(2\chi)k\right)+\frac{\sigma}{\sqrt{2}A}\left(\mu_0+k\mu_4\right)
\end{align}
In this case, these two parts are decoupled from the equations of the $i$- and $j$-imaginary parts of \eqref{bb}. One can solve at each time instant $t$ the equations \eqref{dwew11} in the two variables $A$ and $\chi$, and then in a second step make use of the $i$- and $j$-imaginary parts of \eqref{bb} to solve for $n\theta_0$ and $\varphi$. Because the system is symmetric and ergodic, as long as the noise intensity $\sigma\neq0$, the steady probability density function (PDF) of $n\theta_0$ is uniform. We leave to future investigations the study of the PDF of $\varphi$, and focus on \eqref{dwew11} next.
We multiply both sides of \eqref{dwew11} by $A$ and obtain
\begin{align}
\label{eqdede45}
A'-A\chi' k = \nu A-\frac{3\kappa}{16}A^3-\frac{3\kappa}{32}A^3\cos(2\chi)e^{2\chi k}+\frac{\sigma^2}{4A}\left(1+\tan(2\chi)k\right)+\frac{\sigma}{\sqrt{2}}\left(\mu_0+k\mu_4\right)
\end{align}
We finally split \eqref{eqdede45} into real and $k$-imaginary parts:
\begin{subequations}
\label{eqdede0}
\begin{align}
\label{eqdede}
A' &= \overbrace{\nu A-\frac{3\kappa}{16}A^3-\frac{3\kappa}{32}A^3\cos^2(2\chi)+\frac{\sigma^2}{4A}}^{v_A}+\frac{\sigma}{\sqrt{2}}\mu_0\\
\label{e2}
\chi'&=\overbrace{\frac{3\kappa}{64}A^2\sin(4\chi)-\frac{\sigma^2}{4A^2}\tan(2\chi)}^{v_\chi}+\frac{\sigma}{\sqrt{2}A}\mu_4
\end{align}
\end{subequations}
where we further manipulated the second equation. In both equations, the second to last term on the right hand side arises because the system is stochastic with a non-zero noise intensity $\sigma$, as described in \S\ref{secss}. These terms guarantee that the variables $(A,\chi)$ stay in the bounded domain $\mathcal{R}^+\times (-\pi/4,\pi/4)$ regardless of the noise sources $\mu_0$ and $\mu_4$.

The equations \eqref{eqdede0} describe the evolution of the state variables $(A,\chi)$ as function of time. We study next what happens to the system at different levels $\sigma$ of background noise. For this purpose, we fix the value of the growth rate $\nu/\omega_0$ to $0.04$.

\begin{figure}
\includegraphics[width=\textwidth]{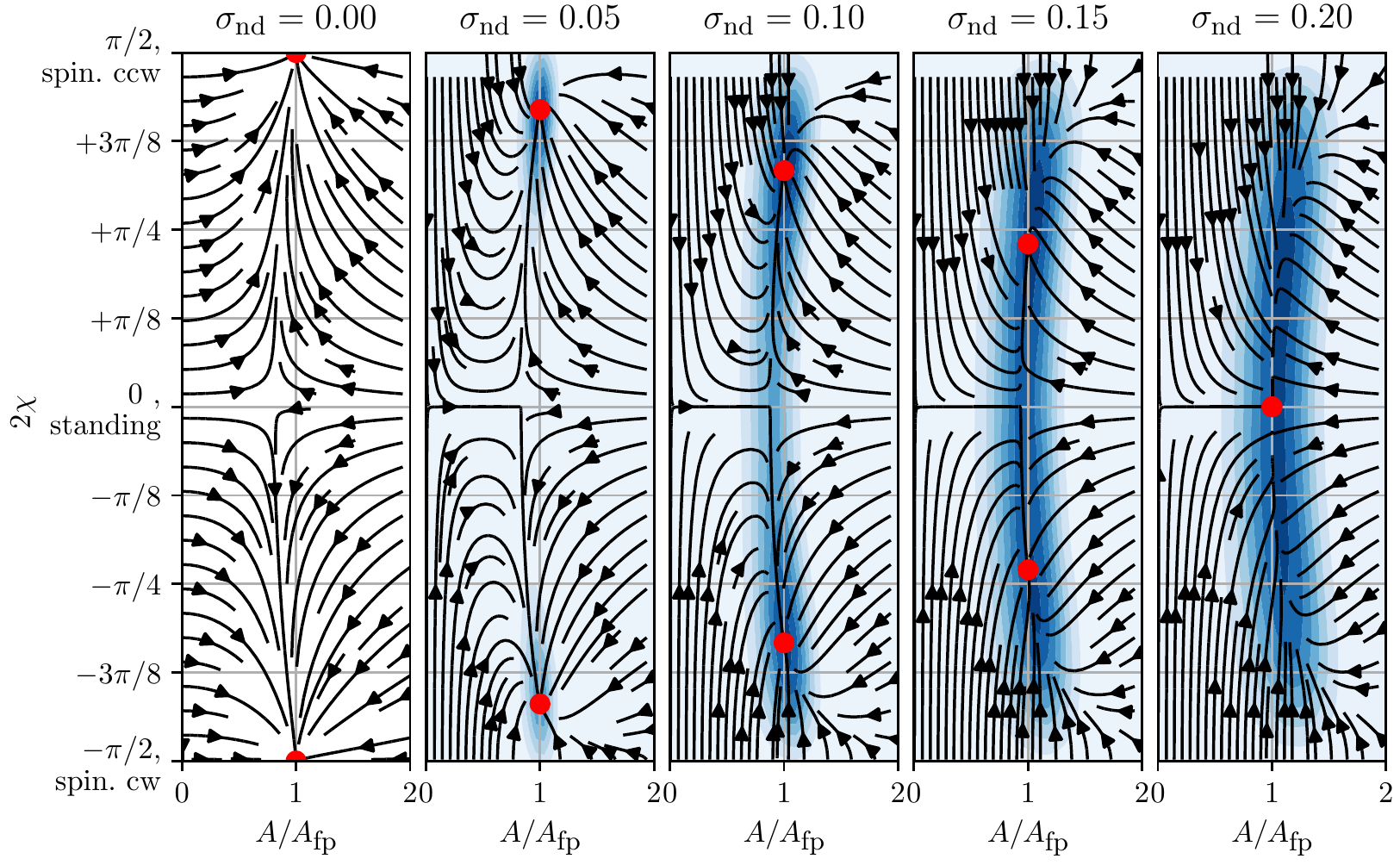}
\caption{Vector field (arrows) and probability density function (PDF, color) of the system state for increasing values of the nondimensional background noise $\sigma_\text{nd}=\sigma/(A_{\text{fp}}\sqrt{\omega_0})$ from left to right, as specified in the title of each frame. The black arrows describe the deterministic vector field on the right hand side of \eqref{eqdede}. The stable fixed points of the vector field are reported with a red dot, and move from a perfect spinning solution ($2\chi=\pm\pi/2$) towards a mixed mode for intermediate levels of noise, and towards a standing solution ($2\chi=0$) at high noise amplitudes. In the first frame on the left, the noise intensity is zero and the deterministic dissipative system converges to the red point after an initial transient. In all subsequent frames the system is stochastic and the state keeps changing with time. In these cases we overlay on the vector field the PDF of the system state from numerical simulations, presented with the colour. The peak of the PDF is close to the stable fixed point. The vector field and the PDF are mirror symmetric around the line $\chi=0$, despite the streamlines not being exactly symmetric. The amplitude $A_{\text{fp}}$ is the amplitude of the stable red fixed point.
\label{fignumnum}
}
\end{figure}

We discuss next the two-dimensional vector field $(v_A,v_\chi)$, i.e. the right hand sides of \eqref{eqdede0} except the last random term proportional to $\sigma/\sqrt{2}$, as the background noise intensity is varied. To this aim, in the first frame of Fig. \ref{fignumnum} we consider the case with $\sigma=0$ and draw with black arrows the streamlines of the state of the system, i.e.~the trajectories $(A(t),\chi(t))$ as function of time starting from a set of points collocated on an equispaced grid. In this deterministic case, after an initial transient, the state of the system converges to either of the stable spinning solutions marked with red fixed points at amplitude $A=A_{\text{fp}}$.

We then increase the level of background noise $\sigma$ from zero, in four steps, in terms of the nondimensional noise intensity $\sigma/(A_{\text{fp}}\sqrt{\omega_0})$. The value of the nondimensional noise intensity is presented in the title of each frame of Fig.~\ref{fignumnum}. Because the system is stochastic, the system state $(A(t),\chi(t))$ does not converge to the red stable fixed point. However, the probability density function (PDF) $P(A,\chi)$ of the system state converges to a steady solution, which is presented in colour from white (zero value) to dark blue (maximum value in each frame). The PDF of each frame is obtained from running a single stochastic simulation of the original equations \eqref{midwife4} for approximately $100'000$ limit cycles, exploiting ergodicity and stationarity of the system\footnote{the slight asymmetries with respect to the axis $\chi=0$ of mirror symmetry are due to the slow convergence of the PDF}. The peaks of $P(A,\chi)$, occurring where the blue is darkest, are very close to the red fixed points. As noise is increased from left to right, the position of the peaks of $P(A,\chi)$ and of the red fixed points move from spinning states towards standing states. We also observe from left to right how the large values of the noise intensity make $P(A,\chi)$ broader, as expected for a system subject to additive noise. This broadening of the PDF is larger in the vertical direction of the nature angle $\chi$ than on the horizontal direction of the amplitude $A$.

\section{Comparison to experiments}
\label{sexper}
In this section, we compare the prediction of the last section with experimental results of nominally rotationally symmetric combustors for which a discussion of the standing and spinning nature of the system is available. Notice however that the equations presented describe also how the system may converge to mixed spinning/standing states because of a loss of rotational symmetry, i.e.~explicit symmetry breaking. This may occur for example because of the introduction of baffles \cite{Dawson2014} or dampers \cite{Stow2003a,Camporeale2004,Lepers2005,Dupere2005,Bothien2015,Zahn2016a,Mensah2016a,Ghirardo2017JFM,Mazur2018a,Yang2019}.
There are currently no experiments that only change the level of background noise $\sigma$ and keep all other relevant parameters the same (one can in the future force an annular system with increasing white noise intensity employing loudspeakers).%
 We can, however, discuss if experiments of annular combustors are noisy or not and if they manifest a preference for standing or spinning states.

To start with, we review the case of combustors exhibiting a low level of background noise. To the knowledge of the authors, this regards only the MICCA combustor equipped with matrix burners \citep{Bourgouin2015}. This experiment exhibits a low level of background noise, based on the fact that the system can converge to spinning or standing states and lingers in the vicinity of these solutions with very tiny variations. We neglect the observed states of the slanted mode \cite{Bourgouin2014_slanted}, which can be explained as a higher-order degeneracy \cite{Orchini2018, Moeck2018, Yang2019a}, which goes beyond the scope of this paper. Also, other dynamical states, e.g.~chaotic solutions, are not discussed. In this case, one can assume that the effect of the background noise is negligible and treat the problem in a deterministic setting, i.e.~by setting $\sigma$ to zero in the equations here discussed. One proves that, neglecting the effect of the azimuthal velocity on the flame response, both standing and spinning solutions can be attractors \citep{Ghirardo16JFM}, depending on the flame saturation.

Theoretical conditions for stable limit-cycle solutions in a deterministic framework have been proposed by \cite{Ghirardo16JFM}, ignoring the effect of transverse forcing on the flames. They have been validated on experimental results of the MICCA combustor in this configuration for spinning modes in the same reference \cite{Ghirardo16JFM}, and for standing modes by Laera et al.~\cite{Laera2017a}. Spinning solutions are predicted to always exist and be stable limit cycles, while standing limit-cycles do not always exist, and when they exist they are stable only if the $N_{2n}$ criterion is respected \cite{Ghirardo16JFM}. This allows for hysteresis and bistability, as observed in experiments \cite{Lepers2005, Prieur2016a}. The key element for the stability of these modes is the term discussed in \S\ref{sechrr2}, from which the $N_{2n}$ criterion arises. In \eqref{termQ}, it is the describing function $Q$, and in particular the structure of its projection on the second harmonic $e^{2in(\theta-\theta_0)}$ of the spatial azimuthal structure of the mode, which governs whether standing modes are stable or not.

We turn our attention to the case of combustors where the background noise $\sigma$ is not negligible. In the annular rig of Worth and Dawson ~\cite[Fig.~8]{Worth2013modaldyn}, \cite[Fig.~2]{Worth2016}, \cite[Fig.~5]{Mazur2018}, \cite[Fig.~2]{Nygard2018}, 
modes are never purely spinning, but always between spinning and standing states, or dominantly standing, depending on the specific configuration and operating condition considered. We argue that the level of the background noise of this combustor is not negligible because the system keeps switching between states in a stochastic manner. In the MICCA combustor equipped with swirl-stabilized flames \cite[Fig.~14]{Bourgouin2013_asme}, the system state is dominantly standing, with a slight preference for mixed states between standing and spinning counter-clockwise. Also here the level of background noise is not negligible, because the PDF of the spin ratio $s$ in their Fig.~14 is heavy-tailed. The spin ratio can be defined as a monotonic function of $\chi$ as $s=\arctan\chi$,  and was modified by \cite{Bourgouin2013_asme} on the definition of \cite{Evesque2003} to characterize the nature of azimuthal instabilities. It serves the same purpose of the nature angle $\chi$.  

The MICCA combustor equipped with liquid spray flames shows strongly standing states \cite[Fig.~8]{Prieur2017}. This experiment should be however considered with caution, due to the occurrence of a blow-off phenomenon that occurs at a much slower timescale and that governs the system dynamics.
The industrial annular combustor discussed by Ghirardo and Bothien \cite[Fig. 6]{Ghirardo2018PRF}, and here reported in Fig.\ref{FigSphere}.c, shows a system state that is never fully spinning. Also this combustor shows a non-negligible level of noise, as depicted by the trajectories of the system state on the Poincar\'e sphere in Fig.~\ref{FigSphere}.b.

All this evidence shows that nominally symmetric annular combustors exhibiting non-negligible levels of background noise are never exactly on a spinning state. This is consistent with the theory presented in \S\ref{secchinoise}, which shows that the noise pushes the system away from spinning solutions and towards standing solutions.

Since the first submission of this manuscript, other works have been published on the subject. \cite{Abel20a} has derived a similar set of equations to \cite{Ghirardo19a} and presented here in \eqref{bb}, albeit under the more restrictive assumption of a cubic flame response and no acoustic coupling with the plenum. They consider how several variations of the system parameter affect the system dynamics of their simplified model, considering also a mean azimuthal flow. In the numerical example of \S\ref{fignumnum} we observe that as the nondimensional noise intensity $\sigma_\text{nd}$ grows, the two red spinning attractors get closer, and the probability of the system state being between the two grows, as recently argued numerically by \cite[Fig.~7]{FaureBeaulieu20b}, by further simplifying equations \eqref{eqdede0} and using a different definition of nondimensional noise intensity. Other works have focused on the identification of the equations proposed in this paper \cite{Ghirardo20ASME} and on the modelling of the flame response and its effects on the dynamics \cite{Ghirardo20combsymp}. A recent work \cite{Mazur2020} considers an azimuthal instability of order $n=1$ that is dominantly spinning in one direction. The authors present also for the first time how the first spatial harmonic ($n=2$) of the same instability travels systematically in the opposite direction, fact that shall be considered to further validate this theory.

\section{Conclusions}
\label{sconcl}
We make use of a novel ansatz in the governing equations of acoustic azimuthal instabilities. We apply the method of averaging and obtain a differential equation governing the amplitude $A$, the orientation angle $n\theta_0$, the nature angle $\chi$ and the temporal phase $\varphi$ of the azimuthal instability. These quantities offer a straightforward physical interpretation, and the resulting equation can be used to study the effect of explicit and spontaneous symmetry breaking in the linear and nonlinear regime, with and without noise. The equations capture for example the effect of a loss of rotational symmetry and the effect of arbitrary nonlinear, time-invariant flame models that depend on acoustic pressure and azimuthal acoustic velocity too. We draw links to the case of hydrodynamic instabilities behind a turbulent rotationally symmetric wake, showing that the same equations apply also in that case.

We then focus on the effect of the background noise and show that it pushes the system away from spinning solutions, towards standing solutions. We present a numerical example where we consider increasing values of the intensity $\sigma$ of the background noise. As $\sigma$  increases, the system nature angle, describing whether the state is standing or spinning, steps away from spinning states towards standing states, consistently with the prediction. 
We show how this prediction is consistent also with existing experimental results. In particular, only experiments subject to a negligible level of noise exhibit purely spinning solutions. Conversely, experiments that are noisy never experience purely spinning solutions, but either states between spinning and standing solutions or mostly standing solutions.

The presented equations apply both to rotationally symmetric and slightly non-symmetric configurations in absence of mean azimuthal flow. Future works can investigate the effect of both explicit and spontaneous symmetry breaking on the solution, in the deterministic or stochastic case. The inclusion in the model of a noise source that depends on the azimuthal coordinate may also be of use in applications.

\appendix

\section{Derivation of the equivalent 1D equations}
\label{apptrivial}
This appendix discusses how to embed, in a one-dimensional equivalent differential equation in the azimuthal direction $\theta$ and time $t$, the effects of the axial dimension of the problem. One can then tackle the equivalent problem, as discussed in the main text of this paper, which has the advantage of having just one spatial dimension.

Apart for some manipulations that are specific to the mapping from a two-dimensional to a one-dimensional periodic problem, the key techniques employed in this appendix are separation of variables and a standard Galerkin series expansion of the solution.

We define the problem in \S\ref{sapp1}, discuss the ansatz in \S\ref{sapp2}, introduce the Galerkin basis in \S\ref{seGal}, project the governing equations on it in \S\ref{secProj}, truncate the series expansion of the solution to the leading term in \S\ref{nm1}, introduce some terms in \S\ref{secspatavg} and derive the final equations in \S\ref{partial}.
\subsection{Problem definition}
\label{sapp1}
In this section we start from the governing equations \eqref{fe2f2e2}, here reported for convenience, which apply in the combustor domain $\Omega=[0,2\pi) \times [0, L]$ with spatial variable ${\bm x} = (\xcyl,\theta)$:
\begin{subequations} 
\label{fe2f2e2c}
\begin{align}
\label{eq_1bc}
\frac{1}{\gamma p_0}\frac{\partial p_1}{\partial t} + \nabla\cdot{\bm{u}}_1&-\frac{\gamma-1}{\gamma p_0}q_1 -\frac{\sigma}{\gamma p_0}{q_s}=0\\
\label{eq_2bc}
\rho_0(\xcyl)\frac{\partial {\bm{u}}_1}{\partial t}+\nabla p_1&=0
\end{align}
\end{subequations}
where \eqref{eq_1} was divided by $\rho_0c^2$ and $\rho_0 c^2=\gamma p_0$ was substituted, with $p_0$ the steady mean pressure that is homogeneous in the domain.
The solution variables are the acoustic field variables $p_1$ and the two-dimensional velocity field ${\bm u}_1$, both depending on the axial coordinate $\xcyl$, on the azimuthal coordinate $\theta$ and on the time $t$. We consider the usual case where the density $\rho_0$ and speed of sound $c$ depend on the axial variable $x$ only. The acoustic boundary conditions, which apply on $\partial\Omega$, are:
\begin{align}
\label{bcadj}
{\bm u}_1\cdot{\bm n}&=\tilde Y\frac{p_1}{\rho_0c}\qquad\text{in}\,\,\partial\Omega
\end{align}
where $\tilde Y$ is the nondimensional acoustic admittance, assumed real-valued,
and ${\bm n}$ is the versor normal to the boundary and pointing outwards from the inside of the domain $\Omega$ to the outside, parallel to the versor $\overrightarrow{\xcyl}$ of the axial component. The boundary $\partial\Omega$ consists of the two circles $\theta\in[0,2\pi)$ at ${\xcyl=0}$ and at ${\xcyl=L}$.
\subsection{Proposed ansatz}
\label{sapp2}
The ansatz \eqref{adep} is then substituted into the equations \eqref{fe2f2e2} to obtain the one-dimensional equations \eqref{dsdqw}. We first substitute \eqref{adep} into \eqref{eq_2bc}:
\begin{align}
\nonumber
\frac{\partial {\bm{u}}_1(\xcyl,\theta,t)}{\partial t}&=-\frac{1}{\rho_0(\xcyl)}\nabla p_1\\
\nonumber
{\bm{u}}_1(\xcyl,\theta,t)&=-\frac{1}{\rho_0(\xcyl)}\nabla\left[\psi(\xcyl) \int^t p(\theta,t)dt\right]\\
\label{e3}
&=-\frac{1}{\rho_0(\xcyl)}\frac{\partial \psi(\xcyl)}{\partial \xcyl}\int^t p(\theta,t)dt\overrightarrow{\xcyl} -\frac{\psi(\xcyl)}{\rho_0(\xcyl)}\int^t \frac{2}{D} \frac{\partial p(\theta,t)}{\partial \theta}dt\overrightarrow{\theta}
\end{align}
where we approximate the derivative in the azimuthal direction $1/\rcyl(\partial/\partial\theta)$ as constant in the radial direction, by approximating $1/\rcyl\approx2/D$ where $D$ is the mean diameter of the annulus. We choose the following ansatz for the acoustic velocity field:
\begin{subequations}
\begin{align}
\label{d1}
{\bm{u}}_1(\xcyl,\theta,t) &= \frac{D\rho_0(\xcyl_b)}{2\rho_0(\xcyl)}\nabla\left[\psi(\xcyl) \int^\theta u(\theta,t)d\theta\right]\\
\label{ytty56}
&= \frac{D\rho_0(\xcyl_b)}{2\rho_0(\xcyl)}\frac{\partial \psi(\xcyl)}{\partial \xcyl} \int^\theta u(\theta,t)d\theta \overrightarrow{\xcyl}+\frac{\rho_0(\xcyl_b)\psi(\xcyl)}{\rho_0(\xcyl)} u(\theta,t)\overrightarrow{\theta}
\end{align}
\end{subequations}
where $\overrightarrow{\theta}$ denotes the versor of the azimuthal component. The structure \eqref{d1} is chosen so that the azimuthal acoustic velocity just downstream of the burners is $u(\theta,t)$. In fact, by evaluating \eqref{ytty56} just downstream of the burners at $\xcyl=\xcyl_b$ where $\psi(\xcyl_b)=1$, we obtain
\begin{align}
\label{e4}
{\bm{u}}_1(\xcyl=\xcyl_b,\theta,t) &= \frac{D}{2}\frac{\partial \psi(\xcyl_b)}{\partial \xcyl} \int^\theta u(\theta,t)d\theta \overrightarrow{\xcyl}+ u(\theta,t)\overrightarrow{\theta}
\end{align}
By comparing either the axial or azimuthal components of \eqref{ytty56} with the respective components of \eqref{e3} we obtain
\begin{align}
\label{df123}
\rho_0(\xcyl_b)\frac{D}{2}\int^\theta u(\theta,t)d\theta &= -\int^tp(\theta,t)dt\\
\label{df123ewew}
\frac{D\rho_0(\xcyl_b)}{2}\frac{\partial u(\theta,t)}{\partial t} + \frac{\partial p(\theta,t)}{\partial \theta}&=0
\end{align}
where in \eqref{df123ewew} we derived both sides by the time $t$ and by the azimuthal coordinate $\theta$. Equation \eqref{df123ewew} can be rewritten in terms of the rescaled acoustic velocity defined in \eqref{def_u1t}:
\begin{align}
\label{fin0gialdas}
\frac{\partial \tilde u(\theta,t)}{\partial t} + \frac{1}{n}\frac{\partial p(\theta,t)}{\partial \theta}&=0
\end{align}

Before proceeding further, we substitute into \eqref{bcadj} the expression of $p_1$ from \eqref{adep} and ${\bm u}_1\cdot{\bm n}$ as the axial component of ${\bm u}_1$ from \eqref{e3}:
\begin{align}
\label{r4r4}
\frac{1}{\rho_0(\xcyl)}\frac{\partial \psi(\xcyl)}{\partial \xcyl}\int^t p(\theta,t)dt = -\frac{Y}{\rho_0c}\psi(\xcyl)p(\theta,t)\quad\text{in}\,\,\partial\Omega
\end{align}
where the final expression has been multiplied by $-1$, and we introduced for convenience the scalar
\begin{align}
\label{DefY}
Y(x)=\begin{cases}
+\tilde Y\qquad \text{if } \,\,x=0\\
-\tilde Y\qquad \text{if } \,\,x=L
\end{cases}
\end{align}
to account for the change of sign of ${\bm u}_1\cdot{\bm n}$ in \eqref{bcadj} at the two extremes of the domain.
The boundary conditions \eqref{r4r4} characterize the one-dimensional acoustic field and will be needed later. The divergence of the velocity field ${\bm{u}}_1$ from \eqref{ytty56} evaluates to:
\begin{align}
\nonumber
\nabla\cdot{\bm{u}}_1 &=\left[\frac{\partial}{\partial \xcyl}\overrightarrow{\xcyl}+\frac{2}{D}\frac{\partial }{\partial \theta}\overrightarrow{\theta}\right]\cdot\left[\frac{D\rho_0(\xcyl_b)}{2\rho_0(\xcyl)}\frac{\partial \psi(\xcyl)}{\partial \xcyl} \int^\theta u(\theta,t)d\theta \overrightarrow{\xcyl}+\frac{\rho_0(\xcyl_b)\psi(\xcyl)}{\rho_0(\xcyl)} u(\theta,t)\overrightarrow{\theta}
\right]\\
\label{eq_e23}
&=-\frac{\partial}{\partial\xcyl}\left(\frac{1}{\rho_0(\xcyl)}\frac{\partial\psi(\xcyl)}{\partial \xcyl}\right)\int^t p(\theta,t)dt
+\frac{2}{D}\psi(\xcyl)\frac{\rho_0(\xcyl_b)}{\rho_0(\xcyl)}\frac{\partial u(\theta,t)}{\partial \theta}
\end{align}
where also \eqref{df123} was substituted in the last step. One then substitutes \eqref{adep} and \eqref{eq_e23} into \eqref{eq_1bc}:
\begin{align}
\label{eq_1bc2}
\frac{\partial p(\theta,t)}{\partial t}\psi(\xcyl) -\gamma p_0\frac{\partial}{\partial\xcyl}\left(\frac{1}{\rho_0(\xcyl)}\frac{\partial\psi(\xcyl)}{\partial \xcyl}\right)\int^t p(\theta,t)dt +\frac{2\rho_0(\xcyl_b)}{D}c^2(\xcyl)\psi(\xcyl)\frac{\partial u(\theta,t)}{\partial \theta}&=(\gamma-1)q_1 +\sigma{q_s} 
\end{align}
where both sides have been multiplied by $\rho_0(\xcyl)c^2(\xcyl)=\gamma p_0$.
\subsection{Galerkin basis}
\label{seGal}
The mode shape $\psi(x)$ respects the boundary condition \eqref{r4r4} but has not been characterized yet. In this section we study the general exact solution of the problem as a series expansion on a Galerkin basis \cite{Morse1953a,Culick2006}. See also \cite{Laurent2019} for recent developments in the search for an optimal basis for the series expansion. The ansatz \eqref{r4r4} will be just the first term of this series expansion, and we will truncate the series after the first term \cite{Ghirardo2017JFM}. This is sometimes referred to as a one-mode Galerkin expansion.

This means that the mode shape $\psi$ is the leading Galerkin mode in the expansion. We choose as Galerkin basis the set of eigenfunctions of the respective pure acoustic problem, obtained from the thermoacoustic problem \eqref{fe2f2e2c} by setting to zero the acoustic sources and sinks:
\begin{subequations} 
\label{fe2f2e2cGal}
\begin{align}
\label{eq_1bcGal}
\frac{1}{\gamma p_0}\frac{\partial p_1}{\partial t} + \nabla\cdot{\bm{u}}_1&=0\\
\label{eq_2bcGal}
\rho_0(\xcyl)\frac{\partial {\bm{u}}_1}{\partial t}+\nabla p_1&=0
\end{align}
 and applying homogeneous boundary conditions:
\begin{align}
\label{bcadjGal}
{\bm u}_1\cdot{\bm n}&=0\qquad\text{in}\,\,\partial\Omega
\end{align}
\end{subequations}
by taking the divergence of \eqref{eq_2bcGal}, and subtracting it to the time derivative of \eqref{eq_1bcGal}, one obtains the wave equation:
\begin{align}
\label{def13g13tf}
\frac{1}{\gamma p_0}\frac{\partial^2 p_1}{\partial t^2} - \nabla\cdot\left(\frac{1}{\rho_0}\nabla p_1\right)&=0
\end{align}

One looks for a solution by means of the technique of separation of variables with structure $p_1=e^{-in\theta}\psi(x)e^{i\omega t}$. By substituting this into \eqref{def13g13tf} one obtains:
\begin{align}
\label{fewf132r2r3}
\frac{-\omega^2}{\gamma p_0}\psi(x)e^{-in\theta}e^{i\omega t} - \left[\frac{\partial}{\partial x} \overrightarrow{x} + \frac{2}{D}\frac{\partial }{\partial \theta}\overrightarrow{\theta}\right]\cdot\left[\left(\frac{1}{\rho_0}\frac{\partial \psi(x)}{\partial x}\overrightarrow{x}-in \frac{2}{D} \frac{\psi(x)}{\rho_0}\overrightarrow{\theta}\right)e^{-in\theta}e^{i\omega t}\right]&=0
\end{align}
One then further simplifies \eqref{fewf132r2r3} and multiplies it by $-\gamma p_0e^{in\theta}e^{-i\omega t}$ to obtain:
\begin{subequations}
\label{feGRQ34WT43}
\begin{align}
\label{fewf132r2r32}
\omega^2\psi(x) + \gamma p_0\frac{\partial}{\partial x} \left(\frac{1}{\rho_0}\frac{\partial \psi(x)}{\partial x}\right) -c^2 \frac{4n^2}{D^2}\psi(x)=0
\end{align}
with boundary conditions
\begin{align}
\label{bcgalerkin}
\frac{\partial \psi}{\partial x} =0\qquad\text{in}\,\,\partial\Omega
\end{align}
\end{subequations}
For the considered azimuthal order $n$, there exists a set of axial acoustic mode shapes $\{\psi_1(x),\psi_2(x), \ldots \}$ with respective eigenfrequencies $\{\omega_1,\omega_2, \ldots \}$ that are solutions of \eqref{feGRQ34WT43}, which are called Galerkin modes. Of all these, we choose the mode shape $\psi$ that appears in the ansatz \eqref{adep} as the one acoustic Galerkin mode that is closest to the thermoacoustic eigenmode of the original problem \eqref{fe2f2e2c} at the amplitude representative of the nonlinear state of the system. These two modes are usually rather similar because the source and sink terms in the thermoacoustic problem are small, and the thermoacoustic mode is a perturbation of the acoustic mode. Under this circumstance, one considers the ansatz \eqref{adep} as a Galerkin series expansion, truncated to the first leading term $\psi$, with a small truncation error. We point out however that there exist some cases, related to intrinsic thermoacoustic modes (ITA), where the acoustic mode and the nonlinear thermoacoustic mode may differ substantially, although this has not yet been quantitatively assessed \cite{Hoeijmakers2014,Courtine2015,Silva2015,Hosseini2018,Orchini2020}.

We can rewrite \eqref{feGRQ34WT43} for the $n$-th eigenmode as:
\begin{subequations}
\begin{align}
\label{fewf132r2r32b}
\gamma p_0\frac{\partial}{\partial x} \left(\frac{1}{\rho_0}\frac{\partial \psi_n(x)}{\partial x}\right)&=-\omega_n^2\psi_n(x)+c^2 \frac{4n^2}{D^2}\psi_n(x)\\
\label{rgrththrst34t}
\frac{\partial \psi_n}{\partial x} &=0\qquad\text{in}\,\,\partial\Omega
\end{align}
\end{subequations}
This expression will be used later.
\subsection{Projection of the governing equations on the Galerkin basis}
\label{secProj}
One then projects the equations on the basis of modes $\{\psi_1,\psi_2,\psi_3, \ldots\}$ that are the eigenfunctions of the acoustic problem. This is reviewed for example in \cite[\S2.3 and \S2.4]{Ghirardo2017JFM}. By truncating the series to the first leading term \eqref{adep}, only the projection on $\psi$ is maintained, and the others on $\psi_1,\psi_2,\ldots$ are discarded. This projection is calculated by multiplying \eqref{eq_1bc2} by the $n$-th mode shape $\psi_n$ and integrating over the whole domain $\Omega$:
\begin{align}
\nonumber
\frac{\partial p(\theta,t)}{\partial t}&\int_\Omega\psi(x)\psi_n(x) dV \overbrace{-\gamma p_0\int_\Omega\frac{\partial}{\partial\xcyl}\left(\frac{1}{\rho_0(\xcyl)}\frac{\partial\psi(\xcyl)}{\partial \xcyl}\right)\psi_n(\xcyl)dV\int^t p(\theta,t)dt}^{R} + \ldots\\
\label{eq_1c3bb}
&\ldots\frac{2\rho_0(\xcyl_b)}{D}\int_\Omega c^2(\xcyl)\psi(\xcyl)\psi_n(\xcyl)dV \frac{\partial u(\theta,t)}{\partial \theta}
=(\gamma-1)\int_\Omega q_1\psi_n(\xcyl)dV +\int_\Omega\sigma{q_s}\psi_n(\xcyl)dV
\end{align}

Integration by parts is applied twice to the second term $R$ in \eqref{eq_1c3bb}:
\begin{align}
\nonumber
R&=-\gamma p_0\int_{\partial\Omega} \frac{1}{\rho_0(\xcyl)}\frac{\partial \psi(\xcyl)}{\partial \xcyl}\psi_n(\xcyl) dS\int^t p(\theta,t)dt
+\gamma p_0\int_\Omega\frac{\partial \psi(\xcyl)}{\partial\xcyl}\frac{1}{\rho_0(\xcyl)}\frac{\partial\psi_n(\xcyl)}{\partial \xcyl}dV\int^t p(\theta,t)dt\\
\nonumber
&=\overbrace{-\gamma p_0\int_{\partial\Omega}\frac{1}{\rho_0(\xcyl)}\frac{\partial \psi(\xcyl)}{\partial \xcyl}\psi_n(\xcyl)dS \int^t p(\theta,t)dt}^{R_1}
+\overbrace{\gamma p_0\int_{\partial\Omega}\frac{1}{\rho_0(\xcyl)}\psi(\xcyl) \frac{\partial \psi_n(\xcyl)}{\partial \xcyl} dS\int^t p(\theta,t)dt\ldots}^{R_2}\\
\label{gg5}
&\qquad\ldots \overbrace{-\gamma p_0\int_\Omega \psi(\xcyl)\frac{\partial}{\partial \xcyl}\left(\frac{1}{\rho_0(\xcyl)}\frac{\partial \psi_n(\xcyl)}{\partial \xcyl}\right)dV\int^t p(\theta,t)dt}^{R_3}
\end{align}
The first integral $R_1$ in \eqref{gg5} evaluates to:
\begin{align}
\label{dasdas13r3}
R_1=-\gamma p_0\int_{\partial\Omega}\frac{1}{\rho_0(\xcyl)}\frac{\partial \psi(\xcyl)}{\partial \xcyl}\psi_n(\xcyl)\,dS\int^t p(\theta,t)dt&=\int_{\partial\Omega}Yc(\xcyl)\psi(\xcyl)\psi_n(\xcyl)dS\,p(\theta,t)
\end{align}
where the boundary condition \eqref{r4r4} was substituted. The second integral $R_2$ in\eqref{gg5} is zero because of the boundary conditions \eqref{rgrththrst34t}. We substitute \eqref{fewf132r2r32b} into the third integral $R_3$ in \eqref{gg5}:
\begin{align}
\label{eqeq00bb}%
R_3 &=\int_\Omega \left(\omega_n^2\psi_n(x)-c^2 \frac{4n^2}{D^2}\psi_n(x)\right)\psi(x)\,dV\int^t p(\theta,t)dt%
\end{align}
Substituting \eqref{dasdas13r3} and \eqref{eqeq00bb} into \eqref{gg5} we obtain:
\begin{align}
\label{fewnkfwe}
\hspace{-.5cm}R = Y\int_{\partial\Omega}c(\xcyl)\psi(\xcyl)\psi_n(\xcyl)dSp(\theta,t) + \left(\omega_n^2\int_\Omega \psi_n(x)\psi(x)\,dV- \int_\Omega c^2(x) \frac{4n^2}{D^2}\psi_n(x)\psi(x)\,dV\right)\int^t p(\theta,t)dt
\end{align}
We substitute \eqref{fewnkfwe} into \eqref{eq_1c3bb}:
\begin{align}
\nonumber
\frac{\partial p(\theta,t)}{\partial t}&\int_\Omega\psi(x)\psi_n(x) dV +\int_{\partial\Omega}Yc(\xcyl)\psi(\xcyl)\psi_n(\xcyl)dS\,p(\theta,t) + \omega_n^2\int_\Omega\psi(x)\psi_n(x)dV\int^t p(\theta,t)dt\,\ldots\\
\nonumber
&\ldots\,- \frac{4n^2}{D^2}\int_\Omega c^2(x)\psi_n(x)\psi(x)dV\int^t p(\theta,t)dt + \frac{2\rho_0(\xcyl_b)}{D}\int_\Omega c^2(\xcyl)\psi(\xcyl)\psi_n(\xcyl)dV\,\frac{\partial u(\theta,t)}{\partial \theta}
=\,\ldots\\
\label{eq_1c3bbc}
&(\gamma-1)\int_\Omega q_1\psi_n(\xcyl)dV +\int_\Omega\sigma{q_s}\psi_n(\xcyl)dV
\end{align}
Equation \eqref{eq_1c3bbc} is the projection of the governing equations on each of the eigenmodes $\psi_n$ for a fixed $n$.
\subsection{Truncation of the solution to the leading term}
\label{nm1}
One then chooses a solution for $p_1$ as a series expansions on the Galerkin basis with the following structure:
\begin{align}
\label{sega}
p_1&=\sum_{s=1}^\infty p^{(s)}(\theta, t)\psi_s(x)
\end{align}
and re-derives the expression for the velocity field ${\bm u}_1$ as in \S\ref{sapp2}, and the expression for the projection of the governing equations on the Galerkin basis as in \S\ref{secProj}. This leads to an equation for each of the terms in the series expansion \eqref{sega}. An example of this new set of equations is presented in \cite[eq. (2.22)]{Ghirardo2017JFM} for a similar case. In the case at hand, we truncate the series \eqref{sega} to just the leading term, which we denote as
\begin{align}
\label{sega2}
p_1&\approx p(\theta, t)\psi(x)
\end{align}
as in the main text in \eqref{adep}, for which \eqref{eq_1c3bbc} holds. We also denote simply with $\omega_0$ its corresponding eigenfrequency. Of all the projections described by \eqref{eq_1c3bbc}, we consider the one on the same mode $\psi_n=\psi$, and introduce
\begin{align}
\label{defLambda}
\Lambda \coloneqq\int_\Omega\psi^2(x) dV
\end{align}
We then substitute $\psi_n=\psi$ and $\omega_n=\omega_0$ into \eqref{eq_1c3bbc} and divide both sides by $\Lambda$:
\begin{align}
\nonumber
\frac{\partial p(\theta,t)}{\partial t}& +\frac{1}{\Lambda}\int_{\partial\Omega}Yc(\xcyl)\psi^2(\xcyl)dS\,p(\theta,t) + \omega_0^2\int^t p(\theta,t)dt- \frac{1}{\Lambda}\frac{4n^2}{D^2}\int_\Omega c^2(x)\psi^2(x)dV\int^t p(\theta,t)dt+\ldots\\
\label{eq_1c3bbcw}
&\ldots\frac{2\rho_0(\xcyl_b)}{D\Lambda}\int_\Omega c^2(x)(\xcyl)\psi^2(\xcyl)dV \frac{\partial u(\theta,t)}{\partial \theta}=\frac{\gamma-1}{\Lambda}\int_\Omega q_1\psi(\xcyl)dV +\frac{\sigma}{\Lambda}\int_\Omega{q_s}\psi(\xcyl)dV
\end{align}
We introduce some spatially averaged quantities in the next section, to then substitute them into \eqref{eq_1c3bbcw} in \S\ref{partial}.
\subsection{Introduction of spatially averaged quantities}
\label{secspatavg}
We introduce the equivalent acoustic damping coefficient
\begin{align}
\label{DefAlpha}
\alpha\coloneqq \frac{1}{\Lambda}  \int_{\partial\Omega}Yc(\xcyl)\psi^2(\xcyl)dS
\end{align}
that appears in the second term on the left hand side of \eqref{eq_1c3bbcw}. We introduce the averaged square of the speed of sound that appears in the fourth and fifth term of \eqref{eq_1c3bbcw}:
\begin{align}
\label{aint2}
\overline{c^2}\coloneqq\frac{1}{\Lambda}\int_\Omega c^2(\xcyl)\psi^2(\xcyl)dV &
\end{align}
On the right hand side of \eqref{eq_1c3bbcw}, in the first term we introduce the projected deterministic flame response:
\begin{align}
\label{Qflame}
q(\theta, p(\theta,t))\coloneqq \frac{1}{\Lambda}
\int_\Omega q_1(\xcyl,\theta, p(\theta,t))\psi(\xcyl)dV 
\end{align}
$q$ is the heat release rate fluctuations on an infinitesimal segment $d\theta$ of the domain, projected on the axial mode shape $\psi$.
In \eqref{Qflame}, the operator $q_1$ has an explicit dependence on the axial coordinate to account for a flame that may not be compact in the axial direction. For a compact flame in a rotationally symmetric combustor, the expression is $q_1(\xcyl,\theta,p(\theta,t))=\delta(\xcyl-\xcyl_b)q_\text{compact}(p(\theta,t))$ and \eqref{Qflame} simplifies to
\begin{align}
q(p(\theta,t)) = \frac{\psi(\xcyl_b)}{\Lambda}q_\text{compact}[p(\theta,t)]
\end{align}
We maintain instead $q$ generic in our case. The last term on the right hand side of \eqref{eq_1c3bbcw} is the projected stochastic flame response:
\begin{align}
\label{aint4}
\frac{\sigma}{\Lambda}\int_\Omega{q_s}(\xcyl,\theta,t)\psi(\xcyl)dV = \frac{\sigma}{\Lambda}\int_\Omega\psi_\xi(\xcyl)\psi(\xcyl)dV\, \xi(\theta,t) = \sigma\xi(\theta,t)
\end{align}
where we assumed without loss of generality that ${q_s}=\psi_\xi(\xcyl)\xi(\theta,t)$ is separable, and $\psi_\xi(\xcyl)$ is scaled such that in the last step $\int_\Omega\psi_\xi(\xcyl)\psi(\xcyl)dV=\Lambda$. Equation \eqref{aint4} serves as a definition of the noise field $\xi(\theta,t)$.
\subsection{Equivalent one-dimensional equation}
\label{partial}
One then substitutes \eqref{DefAlpha}, \eqref{aint2}, \eqref{Qflame} and \eqref{aint4} into \eqref{eq_1c3bbcw}:
\begin{align}
\label{qw2_0}
\frac{\partial p(\theta,t)}{\partial t} & + \alpha p(\theta,t) +  \left(\omega_0^2-\frac{4n^2\overline{c^2}}{D^2}\right)\int^t p(\theta,t)dt + \frac{2\rho_0(\xcyl_b)}{D}\overline{c^2}\frac{\partial u(\theta,t)}{\partial \theta}
=(\gamma-1)q +\sigma\xi
\end{align}
One then substitutes the expression of the acoustic velocity $u$ in terms of the rescaled acoustic velocity $\tilde u=\rho_0(\xcyl_b) D u / 2n$, and $q$ in terms of $\tilde q=(\gamma-1)q -\alpha p$ by means of \eqref{vybunim}:
\begin{align}
\label{qw2}
\frac{\partial p(\theta,t)}{\partial t} & + \left(\omega_0^2-\frac{4n^2\overline{c^2}}{D^2}\right)\int^t p(\theta,t)dt + \frac{4n\overline{c^2}}{D^2}\frac{\partial \tilde u(\theta,t)}{\partial \theta}
=\tilde q +\sigma\xi
\end{align}
The ansatz \eqref{ansatz} for $p$ and $\tilde u$, here reported for convenience,
\begin{subequations} 
\label{ansatzapp}
\begin{align}
\label{ee43431121}
2p(\theta,t) &= e^{-in\theta} \zeta_\text{a}'(t) + \mbox{q.c.} = 2\mbox{Re}\left[e^{-in\theta} \zeta_\text{a}'(t)\right]\\
\label{ee434321121}
2\tilde u(\theta, t) &=ie^{-in\theta} \zeta_\text{a}(t) + \mbox{q.c.}
\end{align}
\end{subequations}
 is then substituted into \eqref{qw2}:
\begin{align}
\label{re234rewb}
e^{-in\theta} \left[\zeta_\text{a}''(t)+\omega_0^2\zeta_\text{a}(t)\right] + \mbox{q.c.} = 2\tilde q[ e^{-in\theta} \zeta_\text{a}'(t) + \mbox{q.c.} ] +2\sigma\xi(\theta,t)
\end{align}
where both terms have been multiplied by 2. Equation \eqref{re234rewb} matches \eqref{midwife} of the main text. Finally, one observes that \eqref{re234rewb} can be obtained in the same manner also from the following one-dimensional equation
\begin{align}
\label{fin2}
\frac{\partial p(\theta,t)}{\partial t} +\omega_0^2 \frac{1}{n}\frac{\partial \tilde u(\theta,t)}{\partial \theta}
=\tilde q +\sigma\xi(\theta,t)
\end{align}
Equations \eqref{fin2} and \eqref{fin0gialdas} are reported in \eqref{dsdqw3} in the main text, and consist of equivalent partial differential equations in $\theta$ and $t$, where the effect of the axial coordinate on the problem is embedded in the equivalent frequency term $\omega_0$ and in the equivalent loss term $-\alpha p$ present in $\tilde q$ on the right hand side of \eqref{fin2}. By substituting back the expressions of $\tilde u $ and $\tilde q$ from \eqref{vybunim} into \eqref{dsdqw3} one recovers \eqref{dsdqw}, which were informally derived in the main text.

\section{Derivation of the averaged equations}
\label{savg1}
In this appendix we discuss the mathematical application of the method of stochastic averaging to the fast ordinary differential equations \eqref{midwife4} to obtain the system of ordinary differential equations \eqref{bb} presented in the paper. No physical interpretation is provided nor additional assumptions are made in this appendix. These materials should be of interest to readers that need to check the results or apply stochastic averaging to similar problems themselves.

In \S\ref{sec_setup} we change variables and express the system dynamics in terms of the slow variables. We then briefly introduce in \S\ref{sec_avg_th} the method of stochastic averaging, specialized for the case of interest. In \S\ref{sec_adapt} we recast our problem to a structure suitable for applying the method. We then calculate some intermediate results in \S\ref{calc1}, \S\ref{calc2} and \S\ref{calc3}. Finally in \S\ref{sectog} all intermediate results lead to the final equation. A nomenclature is provided at the beginning of the paper to ease the reading.
\subsection{Initial setup}
\label{sec_setup}

We choose this ansatz for $\zeta_\text{a}(t)$:
\begin{align}
\label{eqc1}
\zeta_\text{a}(t)&=A(t)e^{in\theta_0(t)}e^{-k\,\chi(t)}e^{j(\omega t + \varphi(t))}/j\omega=-A(t)e^{in\theta_0(t)}e^{-k\,\chi(t)}e^{j(\omega t + \varphi(t))}j/\omega,
\end{align}
which would be the indefinite time integral of \eqref{eqc2} if the four variables $\{A,n\theta_0,\chi,\varphi\}$ did not depend on time. We point out that quaternion numbers are not commutative, so that the $j$ at the end of \eqref{eqc1} cannot be moved to the beginning of the expression. We introduce the complex-valued variables:
\begin{align}
\label{xy}
\begin{cases}
\xapp \coloneqq \zeta_\text{a} -i\zeta_\text{a}i\\
\yapp \coloneqq \zeta_\text{a}'-i\zeta_\text{a}'i
\end{cases}
\end{align}
We recast the second order ordinary differential equation \eqref{midwife4} as two first order equations in terms of $\xapp,\yapp$
\begin{subequations}
\label{bababdas}
\begin{align}
\label{eq1}
\omega \xapp' &= \omega \yapp\\
\label{eq2}
\yapp'+\omega_0^2\xapp&= 2{\qpr} + 2\sigma{\xipr}
\end{align}
\end{subequations}

\subsubsection{Expressions for $\omega \xapp$ and $\yapp$}
First, we substitute \eqref{eqc1} and \eqref{eqc2} into the expressions for $\omega \xapp$ and $\yapp$ defined in \eqref{xy}:
\begin{subequations}
\label{d12}
\begin{align}
\label{d12aaaa}
\omega \xapp =& -Ae^{in\theta_0}e^{-k\,\chi}e^{j(\omega t + \varphi)}j -i\left[-Ae^{in\theta_0}e^{-k\,\chi}e^{j(\omega t + \varphi)}j\right]i\\
\label{d12bbbb}
\yapp=&Ae^{in\theta_0}e^{-k\,\chi}e^{j(\omega t + \varphi)} -i\left[Ae^{in\theta_0}e^{-k\,\chi}e^{j(\omega t + \varphi)}\right]i
\end{align}
\end{subequations}
Quaternions are not commutative, but it is easy to prove by direct substitution that they satisfy these identities:
\begin{align}
\label{eq_ident}
\begin{cases}
e^{i\alpha}i&=ie^{i\alpha}\\
e^{j\alpha}i&=ie^{-j\alpha}\\
e^{k\alpha}i&=ie^{-k\alpha}
\end{cases}\qquad\quad\forall\alpha\in\mathbb{R}
\end{align}
In \eqref{d12aaaa} we first substitute $ji=-k$. Then in both \eqref{d12} we exploit the identities \eqref{eq_ident} to literally move the factor $i$ from the left of the square bracket to the right of the square bracket, to then substitute $ik=-j$ into \eqref{d12aaaa} and $i^2=-1$ into \eqref{d12bbbb}:
\begin{subequations}
\label{d13}
\begin{align}
\label{d13a}
\omega \xapp =& -Ae^{in\theta_0}e^{-k\,\chi}e^{j(\omega t + \varphi)}j + Ae^{in\theta_0}e^{k\,\chi}e^{-j(\omega t + \varphi)}j\\
\label{d13b}
\yapp=&Ae^{in\theta_0}e^{-k\,\chi}e^{j(\omega t + \varphi)}+Ae^{in\theta_0}e^{k\,\chi}e^{-j(\omega t + \varphi)}
\end{align}
\end{subequations}
Equations \eqref{d13} define a change of variables from the two complex-valued fast variables $\{\xapp,\yapp\}$ to the four real-valued slow variables $\{A,n\theta_0,\chi,\varphi\}$. 
\subsubsection{Expressions for $\omega \xapp'$ and $\yapp'$}
We calculate the time derivative of \eqref{d13a} to later substitute it into the left hand side of \eqref{eq1}:
\begin{subequations}
\begin{align}
\label{wx}
\omega \xapp' = e^{in\theta_0}\left(-A'-Ain\theta_0'+Ak\chi'\right)e^{-k\,\chi}e^{j(\omega t + \varphi)}j+Ae^{in\theta_0}e^{-k\,\chi}(\omega+\varphi')e^{j(\omega t + \varphi)} +\Gamma_1 e^{-j(\omega t+\varphi)}
\end{align}
where $\Gamma_1$ consists of all the terms deriving from the second term of \eqref{d13a}, which are all multiplied on the right by the term $e^{-j(\omega t+\varphi)}$. In all the following, terms denoted as $\Gamma_d,\,d=1,2,\ldots,5$ are always multiplied by $e^{-j(\omega t+\varphi)}$ on the right and do not depend directly on the time variable $t$. They are not reported in full length because they will lead to a zero contribution in \S\ref{dasdas1r3}, when the time averaging is carried out on them.

We calculate the time derivative of \eqref{d13b} to later substitute it into the left hand side of \eqref{eq2}:
\begin{align}
\label{ydot}
\yapp'=e^{in\theta_0}\left(A'+Ain\theta_0'-Ak\chi'\right)e^{-k\,\chi}e^{j(\omega t + \varphi)}+Ae^{in\theta_0}e^{-k\,\chi}(\omega+\varphi')e^{j(\omega t + \varphi)}j +\Gamma_2 e^{-j(\omega t+\varphi)}
\end{align}
\end{subequations}

\subsubsection{Equations in terms of the slow variables $\{A,n\theta_0,\chi,\varphi\}$}
We substitute \eqref{wx} and \eqref{d13b} into \eqref{eq1}. We then multiply both sides on the left by $e^{-in\theta_0}$ and on the right by $e^{-j(\omega t + \varphi)}e^{k\,\chi}$. Two terms cancel out and we obtain
\begin{subequations}
\begin{align}
\label{eqkk1}
-\left(A' +Ain\theta_0'-Ak\chi'\right)&e^{-k\chi}je^{k\chi} + A\varphi' + e^{-in\theta_0}\Gamma_3 e^{-2j(\omega t + \varphi)}e^{k\chi}=0
\end{align}
We substitute \eqref{ydot} and \eqref{d13a} into \eqref{eq2}. We then multiply both sides on the left by $e^{-in\theta_0}$ and on the right by $e^{-j(\omega t + \varphi)}e^{k\,\chi}$:
\begin{align}
\nonumber
\left(A' +Ain\theta_0'-Ak\chi'\right) + &A\left(\omega+\varphi'-\frac{\omega_0^2}{\omega}\right)e^{-k\chi}je^{k\chi}+e^{-in\theta_0}\Gamma_4e^{-2j(\omega t + \varphi)}e^{k\chi}=\\
\label{eqkk2}
&+2e^{-in\theta_0}{\qpr}e^{-j(\omega t +\varphi)}e^{k\chi}+2e^{-in\theta_0}\sigma{\xipr} e^{-j(\omega t +\varphi)}e^{k\chi}
\end{align}
\end{subequations}
We now multiply \eqref{eqkk1} on the right by $e^{-k\chi}je^{k\chi}$, sum it to \eqref{eqkk2}, and divide the resulting equation by two:
\begin{align}
\nonumber
\left(A' +Ain\theta_0'-Ak\chi'\right) + &A\left(\frac{\omega}{2}+\varphi'-\frac{\omega_0^2}{2\omega}\right)e^{-k\chi}je^{k\chi}+e^{-in\theta_0}\Gamma_5 e^{-2j(\omega t + \varphi)}e^{k\chi}=\\
\label{pos1}
&+e^{-in\theta_0}{\qpr}e^{-j(\omega t +\varphi)}e^{k\chi}+e^{-in\theta_0}\sigma{\xipr} e^{-j(\omega t +\varphi)}e^{k\chi}
\end{align}
We can divide by $A$ both sides of \eqref{pos1} and obtain
\begin{align}
\label{pos1b}
\frac{A'}{A} +in\theta_0'-k\chi' + &\varphi'e^{-k\chi}je^{k\chi}=f^{(q)} + f^{(\omega)}+ f^{(\Gamma)}+g^{(\xi)}
\end{align}
where we introduce
\begin{subequations}
\label{rhseqeq}
\begin{align}
\label{eqf1a}
f^{(q)}&=\frac{1}{A}e^{-in\theta_0}{\qpr}e^{-j(\omega t +\varphi)}e^{k\chi}\\
\label{eqf1afoorere}
f^{(\omega)}&=\left(-\frac{\omega}{2}+\frac{\omega_0^2}{2\omega}\right)e^{-k\chi}je^{k\chi}\\
\label{eqpp0011}
f^{(\Gamma)}&=-e^{-in\theta_0}\frac{\Gamma_5}{A} e^{-2j(\omega t + \varphi)}e^{k\chi}\\\label{eqf1g1}
g^{(\xi)}&=\frac{\sigma}{A}{\xipr} e^{-j(\omega t +\varphi)}e^{k\chi}
\end{align}
\end{subequations}
where ${\qpr}$ and ${\xipr}$ were defined in \eqref{ert0} and \eqref{ert1} respectively. In this section, by means of a change of variables, we obtained the new equation \eqref{pos1b} in terms of the slow variables $\{A,n\theta_0,\chi,\varphi\}$. To apply the method of stochastic averaging on them, we briefly review the method next.

\subsection{Brief review of the stochastic averaging method}
\label{sec_avg_th}
This section briefly presents known results, reviewed by \cite{Roberts1986}, specialised for the specific case at hand. We consider a system of real valued stochastic differential equations written in the Stratonovich sense:
\begin{align}
\label{eq_2e1ww}
z_v'(t) = \varepsilon^2 f_v(z,t) + \varepsilon g_v(z,t, \xi) \qquad v=0,1,2,3
\end{align}
where $\varepsilon$ is a real valued smallness parameter, $f_v$ is the drift vector and $g_v$ is the diffusion vector:
\begin{align}
\label{strg2}
g_v(z,t,\xi) = g_{vr}(z,t)\xi_r(t) \qquad r=0,1
\end{align}
where we make use of Einstein summation notation and the $\xi_r(t)$ are independent, white Gaussian noise sources with zero expected value. The integer indices $v,r$ can span an arbitrary interval in the general case, and are here already specialized to the problem at hand. The function $f$ and $g$ are periodic in $t$ with period $T$, i.e.~$f(x(t),t)=f(x(t),t+T)$, and similarly for $g$. The symbols $f^{(q)},f^{(\omega)},g^{(\xi)}$ introduced in \eqref{rhseqeq}, consistently  classify the functions in drift terms $f$ and diffusion terms $g$. The resulting system of averaged equations must be interpreted in the It\^{o} sense and are:
\begin{align}
\label{okok}
z_v'(t) = \varepsilon^2 m_v(z) + \varepsilon h_{vb}(z)\mu_b(t)\qquad v=0,1,2,3
\end{align}
where $\mu_b$ are independent, white Gaussian noise sources with unit variance and zero mean, and $m_v$ and $h_{vb}$ are defined as
\begin{subequations}
\label{g0}
\begin{align}
\label{g0a}
m_v(z) &=T^\text{av}_t\left\{f_v(z,t) + \frac{1}{2}\frac{\partial g_{vr}(z,t)}{\partial z_w}g_{wr}(z,t)\right\}\\
\label{g0b}
D_{vw}(z)&= T^\text{av}_t\left\{g_{vr}(z,t)g_{wr}(z,t)\right\}\\
\label{g0b2}
h_{vb}(z)h_{wb}(z)&=D_{vw}(z)\qquad b=0,1,\ldots,B-1
\end{align}
\end{subequations}
where $v,w$ can have value $0,1,2,3$ and $T^\text{av}_t$ is the time averaging operator, defined as:
\begin{align}
\label{eqinreefe}
T^\text{av}_t\left\{f(z(t),t)\right\}=\frac{1}{T}\int_t^{t+T}f(z(t),s)\,ds
\end{align} 
where $f$ is periodic in $t$ as discussed earlier. In the integral \eqref{eqinreefe} the fast dependence of $f$ on time is by means of the integration variable $s$, while the slow dependence of $f$ on time, indirect through the dependence on $z(t)$, is not accounted for in the averaging integral\footnote{indeed $z$ is calculated at time $t$, which is not an integration variable}. In other words, the time averaging is carried out only on the fast timescale, and in the averaging integral the value of the slow variables is frozen.

Notice that the stochastic differential equations \eqref{okok} and \eqref{eq_2e1ww} are not equivalent, and we make use of the same variable $z$ only for ease of notation. One proves however that the stochastic process \eqref{eq_2e1ww} tends to the stochastic process \eqref{okok} in a weakly nonlinear sense \citep{Roberts1986}. From \eqref{g0b}, we observe that the functions $h$ are defined indirectly through the matrix $D$ and the integer $B$, which is the number of noise sources in the averaged system, which will be set later to $4$. For convenience, we rewrite \eqref{g0a} as the sum of two terms:
\begin{subequations}
\label{ben4290342a0}
\begin{align}
\label{ben4290342}
m_v(z) &=m_{v}^{\text{from}\,f}(z)+m_{v}^{\text{from}\,g}(z)\\
\label{eqtmp00a}
m_{v}^{\text{from}\,f}(z)&\coloneqq T^\text{av}_t\left\{f_v(z,t)\right\} \\
\label{eqtmp00b}
m_{v}^{\text{from}\,g}(z)&\coloneqq T^\text{av}_t\left\{\frac{1}{2}\frac{\partial g_{vr}(z,t)}{\partial z_w}g_{wr}(z, t)\right\}
\end{align}
\end{subequations}
We consider next how to apply stochastic averaging to a quaternion-valued case. Consider the quaternion-valued stochastic differential equation:
\begin{align}
\label{eqqqqq}
z'(t) = \varepsilon^2 f(z,t) + \varepsilon g(z,t,\xi)
\end{align}
with $z=z_0+iz_1+jz_2+kz_3$. By substitution we have
\begin{align}
\label{f1wq}
z_0'+iz_1'+jz_2'+kz_3' = \varepsilon^2 f(z,t) + \varepsilon g(z,t,\xi)
\end{align}
We can split \eqref{f1wq} into the real and three imaginary parts and obtain
\begin{subequations}
\label{dew}
\begin{align}
z_0' &=  \mbox{Re}\left[ \varepsilon^2 f(z_0+z_1i+z_2j+z_3,t) + \varepsilon g(z_0+z_1i+z_2j+z_3,t,\xi)\right]\\
z_1' &= \mbox{Im}_i\left[ \varepsilon^2 f(z_0+z_1i+z_2j+z_3,t) + \varepsilon g(z_0+z_1i+z_2j+z_3,t,\xi)\right]\\
z_2' &= \mbox{Im}_j\left[ \varepsilon^2 f(z_0+z_1i+z_2j+z_3,t) + \varepsilon g(z_0+z_1i+z_2j+z_3,t,\xi)\right]\\
z_3' &= \mbox{Im}_k\left[ \varepsilon^2 f(z_0+z_1i+z_2j+z_3,t) + \varepsilon g(z_0+z_1i+z_2j+z_3,t,\xi)\right]
\end{align}
\end{subequations}
Eq. \eqref{dew} is a system of real-valued equations equivalent to \eqref{eqqqqq}. It has the same structure of \eqref{eq_2e1ww} and stochastic averaging can be applied to it.

\subsection{Manipulation of the system to make it amenable of averaging}
\label{sec_adapt}
We rewrite the system \eqref{pos1b} in its real and three imaginary parts, respectively indexed with $0,1,2,3$:
\begin{subequations}
\label{zpo}
\begin{align}
\label{zpo1}
A'/A &= f_0^{(q)}+f_0^{(\omega)}+f_0^{(\Gamma)}+g_0^{(\xi)}\\
\label{zpo2}
n\theta_0'+\varphi'\sin(2\chi) &= f_1^{(q)}+f_1^{(\omega)}+f_1^{(\Gamma)}+g_1^{(\xi)}\\
\label{zpo3}
\varphi'\cos(2\chi) &= f_2^{(q)}+f_2^{(\omega)}+f_2^{(\Gamma)}+g_2^{(\xi)}\\
\label{zpo4}
-\chi' &= f_3^{(q)}+f_3^{(\omega)}+f_3^{(\Gamma)}+g_3^{(\xi)}
\end{align}
\end{subequations}
Eq. \eqref{zpo} does not have the structure \eqref{eqqqqq}, because the left hand sides of \eqref{zpo} are not derivatives of state space variables. 
We multiply \eqref{zpo3} by $\sin(2\chi)$, and subtract it to \eqref{zpo2} multiplied by $\cos(2\chi)$. We obtain:
\begin{align}
\label{zpo2bis}
n\theta_0'\cos(2\chi)&= \left(f_1^{(q)}+f_1^{(\omega)}+f_1^{(\Gamma)}+g_1^{(\xi)}\right)\cos(2\chi)-\left(f_2^{(q)}+f_2^{(\omega)}+f_2^{(\Gamma)}+g_2^{(\xi)}\right)\sin(2\chi)
\end{align}
We divide \eqref{zpo3} and \eqref{zpo2bis} by $\cos(2\chi)$. The system \eqref{zpo} becomes:

\begin{align}
\label{cd}
\begin{cases}
(\ln A)' = f^{(q)}_0+f^{(\omega)}_0+f_0^{(\Gamma)}+g_0^{(\xi)}\\
n\theta_0'= f^{(q)}_1+f^{(\omega)}_1+f_1^{(\Gamma)}-(f^{(q)}_2+f^{(\omega)}_2+f_2^{(\Gamma)})\tan(2\chi)+g_1^{(\xi)} -g_2^{(\xi)}\tan(2\chi)\\
\varphi' = (f^{(q)}_2+f^{(\omega)}_2)/\cos(2\chi)+g_2^{(\xi)}/\cos(2\chi)\\
\chi' = -f^{(q)}_3-f^{(\omega)}_3-g_3^{(\xi)}
\end{cases}
\end{align}
The system \eqref{cd} now matches exactly the structure \eqref{eq_2e1ww} with $\varepsilon=1$, with these definitions:
\begin{subequations}
\label{zssspob}
\begin{align}
&\begin{cases}
z_0&= \ln (A/p_0)\\
z_1&=n\theta_0\\
z_2&=\varphi\\
z_3&=\chi
\end{cases}
&\begin{cases}
f_0&=f^{(q)}_0+f^{(\Gamma)}_0\\
f_1&=(f^{(q)}_1+f^{(\omega)}_1+f^{(\Gamma)}_1)-(f^{(q)}_2+f^{(\omega)}_2+f^{(\Gamma)}_2)\tan(2\chi)\\
f_2&=(f^{(q)}_2+f^{(\omega)}_2+f^{(\Gamma)}_2)/\cos(2\chi)\\
f_3&=-f^{(q)}_3-f^{(\Gamma)}_3
\end{cases}
&\begin{cases}
g_0&=g_0^{(\xi)}\\
g_1&=g_1^{(\xi)} -g_2^{(\xi)}\tan(2\chi)\\
g_2&=g_2^{(\xi)}/\cos(2\chi)\\
g_3&=-g_3^{(\xi)}
\end{cases}
\end{align}
\end{subequations}
where we exploited the fact that $f^{(\omega)}_0=f^{(\omega)}_3=0$. The terms on the right hand sides are defined in \eqref{rhseqeq}. In the definition of $z_0= \ln (A/p_0)$ appears a normalizing factor $p_0$ so that the argument of the logarithm is non-dimensional. Notice however that its time derivative $\ln (A/p_0)' = A'/A$ is idependent of $p_0$.

The equations resulting from stochastic averaging are \eqref{okok}. The terms appearing in the equations can be calculated by means of \eqref{ben4290342a0} and of \eqref{g0b} and \eqref{g0b2}. We calculate next in \S\ref{calc1} the deterministic part of the drift term $m_{v}^{\text{from}\,f}$ defined in \eqref{eqtmp00a}, in \S\ref{calc2} the the stochastic part of the drift term $m_{v}^{\text{from}\,g}$ defined in \eqref{eqtmp00b}, and finally the diffusion matrix $D$ defined in \eqref{g0b2} in \S\ref{calc3}.

\subsection{Calculation of the part of the drift arising from the deterministic terms}
\label{calc1}
We substitute the components of $f$ from \eqref{zssspob} into \eqref{eqtmp00a}:
\begin{subequations}
\begin{align}
\label{pp1}
m_{0}^{\text{from}\,f}&=T^\text{av}_t\left\{f_0^{(q)}\right\}+T^\text{av}_t\left\{f_0^{(\Gamma)}\right\}\\
\nonumber
m_{1}^{\text{from}\,f}&=\left(T^\text{av}_t\left\{f^{(q)}_1\right\}+T^\text{av}_t\left\{f^{(\omega)}_1\right\}+T^\text{av}_t\left\{f^{(\Gamma)}_1\right\}\right)\\
\label{pp2}
&\qquad-\left(T^\text{av}_t\left\{f^{(q)}_2\right\}+T^\text{av}_t\left\{f^{(\omega)}_2\right\}+T^\text{av}_t\left\{f^{(\Gamma)}_2\right\}\right)\tan(2\chi)\\
\label{pp3}
m_{2}^{\text{from}\,f}&=\left(T^\text{av}_t\left\{f^{(q)}_2\right\}+T^\text{av}_t\left\{f^{(\omega)}_2\right\}+T^\text{av}_t\left\{f^{(\Gamma)}_2\right\}\right)/\cos(2\chi)\\
\label{pp4}
m_{3}^{\text{from}\,f}&=-T^\text{av}_t\left\{f^{(q)}_3\right\}-T^\text{av}_t\left\{f^{(\Gamma)}_3\right\}
\end{align}
\end{subequations}
where we exploited that $\chi$ is a slow variable and that the operator $T^\text{av}_t$ is linear. Next, instead of calculating separately the four terms $T^\text{av}_t\left\{f_v^{(q)}\right\}$ for $v=0,1,2,3$, we can make use of the linearity of the operator $T^\text{av}_t$ and write that
\begin{align}
\label{eq_spli5} 
T^\text{av}_t\left\{f_0^{(q)}+if_1^{(q)}+jf_2^{(q)}+kf_3^{(q)}\right\}=T^\text{av}_t\left\{f^{(q)}\right\}
\end{align}
and then consider later the four components, e.g.~we can extract $T^\text{av}_t\left\{f_2^{(q)}\right\}$ as the $j$-th imaginary part of $T^\text{av}_t\left\{f^{(q)}\right\}$ once that has been calculated. We calculate first the average of $f^{(q)}$ in \S\ref{dasdasda}, then the average of $f^{(\omega)}$ in \S\ref{dasdas1r3}, and finally the average of $f^{(\Gamma)}$ in \S\ref{dasdas1r42}.
\subsubsection{Averaging $f^{(q)}$}
\label{dasdasda}
We substitute first \eqref{eqf1a}, and then \eqref{ert0}
\begin{subequations}
\begin{align}
T^\text{av}_t\left\{f^{(q)}\right\}&= T^\text{av}_t\left\{\frac{1}{A}e^{-in\theta_0}{\qpr}e^{-j(\omega t +\varphi)}e^{k\chi}\right\}\\
&=\frac{e^{-in\theta_0}}{A}T^\text{av}_t\left\{\frac{1}{\pi}\int_0^{2\pi}e^{in\theta}\tilde q_1d\theta\,e^{-j\omega t}\right\}e^{-j\varphi}e^{k\chi}\\
\label{e12rr}
&=e^{-in\theta_0}\frac{\omega}{2\pi A}\left\{\int_{0}^{2\pi/\omega} \frac{1}{\pi}\left[\int_0^{2\pi}e^{in\theta}\tilde q_1d\theta\right]e^{-j\omega t}\,dt\right\}e^{-j\varphi}e^{k\chi}
\end{align}
\end{subequations}
We reorder the terms in \eqref{e12rr}  to obtain:
\begin{align}
\label{trg}
T^\text{av}_t\left\{f^{(q)}\right\}=e^{-in\theta_0}\frac{1}{\pi}\int_0^{2\pi} e^{in\theta}\frac{\omega}{2\pi A}\int_{0}^{2\pi/\omega}\tilde q_1[p(t)]e^{-j\omega t}dt\,d\theta\,e^{-j\varphi}e^{k\chi}
\end{align}
We rewrite \eqref{q} as 
\begin{align}\label{dqw1}
p=\alpha\cos\phi + \beta\sin\phi
\end{align}
where
\begin{subequations}
\begin{align}
\label{q1fqw3}
&\begin{cases}
\alpha&\coloneqq A\cos(n\theta-n\theta_0)\cos\chi\\
\beta&\coloneqq A\sin(n\theta-n\theta_0)\sin\chi
\end{cases}\\
\label{eqfastphi}
&\phi=\omega t+\varphi
\end{align}
\end{subequations}
and $\phi$ is the fast varying phase. We introduce
\begin{align}
\label{eqwdqsa}
\begin{cases}
\alpha &= A_p\cos\gamma\\
\beta&=A_p\sin\gamma
\end{cases}
\end{align}
and its inverse transformation
\begin{subequations}
\begin{align}
\label{ampl_local}
A_p &\coloneqq A\sqrt{\cos^2(n\theta-n\theta_0)\cos^2(\chi)+\sin^2(n\theta-n\theta_0)\sin^2(\chi)}\\
\gamma &\coloneqq \mbox{Arg}[\alpha + j\beta] = \mbox{Arg}[\cos(n\theta-n\theta_0)\cos(\chi)+j\sin(n\theta-n\theta_0)\sin(\chi)]
\end{align}
\end{subequations}
where the definition \eqref{ampl_local} is reported also in \eqref{ampl_local2} and a physical interpretation is provided there.
We substitute \eqref{eqwdqsa} into \eqref{dqw1} and obtain this expression for the pressure field:
\begin{align}\label{dqw2}
p=A_p\cos(\omega t+\varphi-\gamma)
\end{align}
We substitute \eqref{dqw2} into \eqref{trg}:
\begin{align}
\label{trg2}
T^\text{av}_t\left\{f^{(q)}\right\}=e^{-in\theta_0}\frac{1}{\pi}\int_0^{2\pi} e^{in\theta}\frac{\omega}{2\pi A}\int_{0}^{2\pi/\omega}\tilde q_1[A_p\cos(\omega t+\varphi-\gamma)]e^{-j\omega t}dt\,d\theta\, e^{-j\varphi}e^{k\chi}
\end{align}
We apply the change of variable $s=t+\varphi/\omega-\gamma/\omega$ to \eqref{trg2}:
\begin{align}
\label{trg3}
T^\text{av}_t\left\{f^{(q)}\right\}=e^{-in\theta_0}\frac{1}{\pi}\int_0^{2\pi} e^{in\theta}\frac{\omega}{2\pi A}\int_{0}^{2\pi/\omega}\tilde q_1[A_p\cos(\omega s)]e^{-j(\omega s -\varphi+\gamma)}ds\,d\theta\,e^{-j\varphi}e^{k\chi}
\end{align}
We split and reorder the terms in \eqref{trg3}:
\begin{align}
\label{trg3a}
T^\text{av}_t\left\{f^{(q)}\right\}=e^{-in\theta_0}\frac{1}{\pi}\int_0^{2\pi} e^{in\theta}\frac{A_p}{2A}\left[\frac{\omega}{A_p\pi}\int_{0}^{2\pi/\omega}\tilde q_1[A_p\cos(\omega s)]e^{-j\omega s}ds\right]e^{-j\gamma}e^{j\varphi} d\theta\,e^{-j\varphi}e^{k\chi}
\end{align}
The term within square brackets in \eqref{trg3a} is the definition \eqref{eqq1pQ} of the describing function $Q_\theta(A_p)$:
\begin{align}
\label{trg3b}
T^\text{av}_t\left\{f^{(q)}\right\}=e^{-in\theta_0}\frac{1}{\pi}\int_0^{2\pi} e^{in\theta}\frac{A_p}{2A}e^{-j\gamma}Q_\theta(A_p) d\theta\,e^{k\chi}
\end{align}
We now observe from \eqref{eqwdqsa} that
\begin{align}
\label{e12re1}
A_pe^{-j\gamma} = \alpha-j\beta
\end{align}
We substitute \eqref{q1fqw3} into \eqref{e12re1}, \eqref{e12re1} into \eqref{trg3b}, and reorder the terms:
\begin{align}
\label{trg4}
T^\text{av}_t\left\{f^{(q)}\right\}=\frac{1}{2}\frac{1}{\pi}\int_0^{2\pi} \underbrace{e^{i(n\theta-n\theta_0)}\left[\cos(n\theta-n\theta_0)\cos\chi -j\sin(n\theta-n\theta_0)\sin\chi\right]}_KQ_\theta(A_p)d\theta e^{k\chi}
\end{align}
We focus on the term $K$, which respects the identity
\begin{align}
\label{id2}
K&=e^{i(n\theta-n\theta_0)}\left[\cos(n\theta-n\theta_0)\cos\chi -j\sin(n\theta-n\theta_0)\sin\chi\right]=\frac{1}{2}e^{2i(n\theta-\theta_0)}e^{k\chi}+\frac{1}{2}e^{-k\chi}
\end{align}
We substitute \eqref{id2} into \eqref{trg4}
\begin{align}
\label{trg5}
T^\text{av}_t\left\{f^{(q)}\right\}=\frac{1}{2}\frac{1}{2\pi}\int_0^{2\pi} \left(e^{i2n(\theta-\theta_0)}e^{k\chi}+e^{-k\chi}\right)Q_\theta(A_p)d\theta\,e^{k\chi}
\end{align}
We keep the result \eqref{trg5} for later.
\subsubsection{Averaging $f^{(\omega)}$}
\label{dasdas1r3}
We substitute the expression for $f^{(\omega)}$ from \eqref{eqf1afoorere}:
\begin{align}
T^\text{av}_t\left\{f^{(\omega)}\right\}&=T^\text{av}_t\left\{\left(-\frac{\omega}{2}+\frac{\omega_0^2}{2\omega}\right)e^{-k\chi}je^{k\chi}\right\}
\end{align}
The argument of the averaging operator does not depend directly on the time $t$, so that trivially
\begin{align}
\label{eq_sec}
T^\text{av}_t\left\{f^{(\omega)}\right\}&=f^{(\omega)}=\left(-\frac{\omega}{2}+\frac{\omega_0^2}{2\omega}\right)e^{-k\chi}je^{k\chi}
\end{align}

\subsubsection{Averaging $f^{(\Gamma)}$}
\label{dasdas1r42}
We substitute the expression for $f^{(\Gamma)}$ from \eqref{eqpp0011}:
\begin{align}
\label{eqlastlast}
T^\text{av}_t\left\{f^{(\Gamma)}\right\}&=T^\text{av}_t\left\{-e^{-in\theta_0}\frac{\Gamma_5}{A} e^{-2j(\omega t + \varphi)}e^{k\chi}\right\}=-e^{-in\theta_0}\frac{\Gamma_5}{A} T^\text{av}_t\left\{e^{-2j(\omega t + \varphi)}\right\}e^{k\chi}=0
\end{align}
where in the first step we take out of the averaging operator all variables that do not directly depend on the time $t$, and in the second step we exploit the fact that, by definition \eqref{eqinreefe} of averaging,
\begin{align}
T^\text{av}_t\left\{e^{-2j(\omega t + \varphi)}\right\}=\frac{\omega}{2\pi}\int_{0}^{2\pi/\omega}e^{-2j(\omega t + \varphi)}dt = 0
\end{align}

\subsection{Calculation of the part of the drift arising from the stochastic terms}
\label{calc2}
In this section we calculate the term \eqref{eqtmp00b}. 
We substitute \eqref{ert1} into \eqref{eqf1g1}:
\begin{align}
\label{g100}
g^{(\xi)}&=\frac{\sigma}{A}\left[\xi_0(t)+i\xi_1(t)\right]e^{-j(\omega t +\varphi)}e^{k\chi}
\end{align}
We explicitly split \eqref{g100} into components:
\begin{subequations}
\begin{align}
\nonumber
g^{(\xi)}=&\frac{\sigma}{A} (\xi_0(t)+i\xi_1(t))\left(\cos(\omega t+\varphi)-j\sin(\omega t+\varphi)\right)\left(\cos\chi+k\sin\chi\right)\\
\nonumber
=&\left(\frac{\sigma}{A}c_{\phi}c_\chi\right)\xi_0+\left(\frac{\sigma}{A}c_{\phi }c_\chi\right)\xi_1i+\left(-\frac{\sigma}{A}s_{\phi }c_\chi\right)\xi_0j+\left(-\frac{\sigma}{A}s_{\phi }c_\chi\right)\xi_1k\\
\label{g101}
+&\left(\frac{\sigma}{A}c_{\phi}s_\chi\right)\xi_0k+\left(-\frac{\sigma}{A}c_{\phi }s_\chi\right)\xi_1j+\left(-\frac{\sigma}{A}s_{\phi }s_\chi\right)\xi_0i+\left(\frac{\sigma}{A}s_{\phi }s_\chi\right)\xi_1\\
\nonumber
=&\left[\left(\frac{\sigma}{A}c_{\phi}c_\chi\right)\xi_0+\left(\frac{\sigma}{A}s_{\phi }s_\chi\right)\xi_1\right]
+\left[\left(-\frac{\sigma}{A}s_{\phi }s_\chi\right)\xi_0+\left(\frac{\sigma}{A}c_{\phi }c_\chi\right)\xi_1\right]i\\
+&\left[\left(-\frac{\sigma}{A}s_{\phi }c_\chi\right)\xi_0+\left(-\frac{\sigma}{A}c_{\phi }s_\chi\right)\xi_1\right]j
+\left[\left(\frac{\sigma}{A}c_{\phi}s_\chi\right)\xi_0+\left(-\frac{\sigma}{A}s_{\phi }c_\chi\right)\xi_1\right]k
\end{align}
\end{subequations}
where we make use of the notation $c_f=\cos(f)$ and $s_f=\sin(f)$ for brevity. The functions $g_v$ for $v=0,1,2,3$ from \eqref{zssspob} become
\begin{align}
\label{Eqwfcqw}
\begin{cases}
g_0&=\left(\frac{\sigma}{A}c_{\phi}c_\chi\right)\xi_0+\left(\frac{\sigma}{A}s_{\phi }s_\chi\right)\xi_1\\
g_1&=\left(-\frac{\sigma}{A}s_{\phi }s_\chi\right)\xi_0+\left(\frac{\sigma}{A}c_{\phi }c_\chi\right)\xi_1-\left[\left(-\frac{\sigma}{A}s_{\phi }c_\chi\right)\xi_0+\left(-\frac{\sigma}{A}c_{\phi }s_\chi\right)\xi_1\right]\tan(2\chi)\\
g_2&=\left[\left(-\frac{\sigma}{A}s_{\phi }c_\chi\right)\xi_0+\left(-\frac{\sigma}{A}c_{\phi }s_\chi\right)\xi_1\right]/\cos(2\chi)\\
g_3&=\left(-\frac{\sigma}{A}c_{\phi}s_\chi\right)\xi_0+\left(\frac{\sigma}{A}s_{\phi }c_\chi\right)\xi_1
\end{cases}
\end{align}
From \eqref{Eqwfcqw} we now express the various terms $g_{vr}$ appearing on the right hand side of \eqref{eqtmp00b}:
\begin{subequations}
\label{eq9090a}
\begin{alignat}{3}
g_{00} &= \frac{\sigma}{A}c_{\phi}c_\chi&\qquad\qquad
g_{01}&=\frac{\sigma}{A}s_{\phi }s_\chi\\
g_{10} &= -\frac{\sigma}{A}s_{\phi }s_\chi+\frac{\sigma}{A}s_\phi c_\chi \tan(2\chi) &\qquad\qquad
g_{11}&=\frac{\sigma}{A}c_{\phi }c_\chi+\frac{\sigma}{A}c_\phi s_\chi\tan(2\chi)\\
g_{20} &= -\frac{\sigma}{A\cos(2\chi)}s_{\phi }c_\chi&\qquad\qquad
g_{21}&=-\frac{\sigma}{A\cos(2\chi)}c_{\phi }s_\chi\\
g_{30} &=- \frac{\sigma}{A}c_{\phi}s_\chi&\qquad\qquad
g_{31}&=+\frac{\sigma}{A}s_{\phi }c_\chi
\end{alignat}
\end{subequations}
We observe that $g_{vr}$ depend only on $A$, $\varphi$ and $\chi$, so that the summation over $w$ in \eqref{eqtmp00b} counts only these three terms:
\begin{align}
\label{eqwewq4433}
m_{v}^{\text{from}\,g}&=T^\text{av}_t\left\{\frac{1}{2}\frac{\partial g_{vr}}{\partial z_w}g_{wr}\right\}
=\frac{1}{2}T^\text{av}_t\left\{\frac{\partial g_{vr}}{\partial \ln A}g_{0r}+\frac{\partial g_{vr}}{\partial \varphi}g_{2r}+\frac{\partial g_{vr}}{\partial \chi}g_{3r}\right\}\qquad v=0,1,2,3
\end{align}
We then expand the summation over $r=0,1$ in \eqref{eqwewq4433}, and express the derivative by $\ln A$ in terms of the derivative by $A$:
\begin{align}
\nonumber
m_{v}^{\text{from}\,g}=\frac{1}{2}T^\text{av}_t\Big\{
&A\frac{\partial g_{v0}}{\partial A}g_{00}+\frac{\partial g_{v0}}{\partial \varphi}g_{20}+\frac{\partial g_{v0}}{\partial \chi}g_{30}\\
\label{Eqwdqwcqs212}
+&A\frac{\partial g_{v1}}{\partial A}g_{01}+\frac{\partial g_{v1}}{\partial \varphi}g_{21}+\frac{\partial g_{v1}}{\partial \chi}g_{31}\Big\}\qquad v=0,1,2,3
\end{align}
We substitute \eqref{eq9090a} into \eqref{Eqwdqwcqs212} and obtain:
\begin{subequations}
\begin{align}
\nonumber
m_{v}^{\text{from}\,g}=\frac{1}{2}T^\text{av}_t\Big\{&A\frac{\partial g_{v0}}{\partial A}\frac{\sigma}{A}c_{\phi}c_\chi
+\frac{\partial g_{v0}}{\partial \varphi}\left(-\frac{\sigma}{A\cos(2\chi)}s_{\phi }c_\chi\right)
+\frac{\partial g_{v0}}{\partial \chi}\left(-\frac{\sigma}{A}c_{\phi}s_\chi\right)\\
\label{Eqwdqwcqs212b}
+&A\frac{\partial g_{v1}}{\partial A}\frac{\sigma}{A}s_{\phi }s_\chi
+\frac{\partial g_{v1}}{\partial \varphi}\left(-\frac{\sigma}{A\cos(2\chi)}c_{\phi }s_\chi\right)
+\frac{\partial g_{v1}}{\partial \chi}\frac{\sigma}{A}s_{\phi }c_\chi\Big\}\\
\nonumber
=\frac{\sigma}{2 A}T^\text{av}_t\Big\{&A\frac{\partial g_{v0}}{\partial A}c_{\phi}c_\chi
-\frac{\partial g_{v0}}{\partial \varphi}\frac{s_{\phi }c_\chi}{\cos(2\chi)}
-\frac{\partial g_{v0}}{\partial \chi}c_{\phi}s_\chi\\
\label{Eqwdqwcqs212c}
+&A\frac{\partial g_{v1}}{\partial A}s_{\phi }s_\chi
-\frac{\partial g_{v1}}{\partial \varphi}\frac{c_{\phi }s_\chi}{\cos(2\chi)}
+\frac{\partial g_{v1}}{\partial \chi}s_{\phi }c_\chi\Big\}
\end{align}
\end{subequations}
For the specific structure of $g_{vr}$ of \eqref{eq9090a} we also observe that 
\begin{subequations}
\label{Fewvw}
\begin{align}
\frac{\partial g_{v0}}{\partial A} &= -\frac{1}{A}g_{v0} \qquad v=0,1,2,3\\
\frac{\partial g_{v1}}{\partial A} &= -\frac{1}{A}g_{v1} \qquad v=0,1,2,3
\end{align}
\end{subequations}
We substitute \eqref{Fewvw} into \eqref{Eqwdqwcqs212c} and obtain:
\begin{align}
\nonumber
m_{v}^{\text{from}\,g}=&\frac{\sigma}{2 A}T^\text{av}_t\Big\{- g_{v0}c_{\phi}c_\chi
-\frac{\partial g_{v0}}{\partial \varphi}\frac{s_{\phi }c_\chi}{\cos(2\chi)}
-\frac{\partial g_{v0}}{\partial \chi}c_{\phi}s_\chi-\\
\label{mj}
& \qquad\qquad g_{v1}s_{\phi }s_\chi
-\frac{\partial g_{v1}}{\partial \varphi}\frac{c_{\phi }s_\chi}{\cos(2\chi)}
+\frac{\partial g_{v1}}{\partial \chi}s_{\phi }c_\chi\Big\}\qquad v=0,1,2,3
\end{align}
Here and in the following, it is important to observe that only the variable $\phi$, appearing in $c_\phi$ and $s_\phi$ in \eqref{mj}, depends directly on the time $t$ (see definition \eqref{eqfastphi}). By direct substitution into \eqref{eqinreefe}, one proves the following identities:
\begin{subequations}
\label{cruc}
\begin{align}
\label{cruc1}
T^\text{av}_t\left\{c_\phi^2\right\} &= T^\text{av}_t\left\{s_\phi^2\right\} = \frac{1}{2}\\
\label{cruc2}
T^\text{av}_t\left\{c_\phi s_\phi\right\} &= 0
\end{align}
\end{subequations}
All the other variables are slow variables of the time $t$, and are constant with regards to the averaging operator, as discussed just after \eqref{eqinreefe}. By making use of the identities \eqref{cruc} the drift functions $m_{v}^{\text{from}\,g}$ are calculated:
\begin{align}
\label{eqm1g}
m_{0}^{\text{from}\,g}&=\frac{\sigma^2}{4A^2}\\
\label{eqm2g}
m_{1}^{\text{from}\,g}&=0\\
\label{m3}
m_{2}^{\text{from}\,g}&=0\\
\label{mqw4}
m_{3}^{\text{from}\,g}&=-\frac{\sigma^2}{4 A^2}\tan(2\chi)
\end{align}

\subsection{Calculation of the diffusion term}
\label{calc3}
We calculate the diffusion matrix $D$ first. We expand the summation in \eqref{g0b} over $r$
\begin{align}
\label{g0c}
D_{vw}(z)&= T^\text{av}_t\left\{g_{v0}(z,t)g_{w0}(z,t)\right\} + T^\text{av}_t\left\{g_{v1}(z,t)g_{w1}(z,t)\right\}\qquad \forall v,w = 0,1,2,3
\end{align}
Also in this section, the identities \eqref{cruc} are used at length to simplify the expressions. We calculate each term individually next. We start with the first row of $D$, from \eqref{g0c}:
\begin{subequations}
\label{djr0}
\begin{align}
\label{g0c11}
D_{00}(z)&= T^\text{av}_t\left\{g_{00}(z,t)g_{00}(z,t)\right\} + T^\text{av}_t\left\{g_{01}(z,t)g_{01}(z,t)\right\}=\frac{\sigma^2}{2A^2}c_\chi^2+\frac{\sigma^2}{2A^2}s_\chi^2=\frac{\sigma^2}{2A^2}\\
\label{g0c12}
D_{01}(z)&= T^\text{av}_t\left\{g_{00}(z,t)g_{10}(z,t)\right\} + T^\text{av}_t\left\{g_{01}(z,t)g_{11}(z,t)\right\}=0\\
\label{g0c13}
D_{02}(z)&= T^\text{av}_t\left\{g_{00}(z,t)g_{20}(z,t)\right\} + T^\text{av}_t\left\{g_{01}(z,t)g_{21}(z,t)\right\}=0\\
\label{g0c14}
D_{03}(z)&= T^\text{av}_t\left\{g_{00}(z,t)g_{30}(z,t)\right\} + T^\text{av}_t\left\{g_{01}(z,t)g_{31}(z,t)\right\}=-\frac{\sigma^2}{2A^2}c_\chi s_\chi+\frac{\sigma^2}{2A^2}c_\chi s_\chi = 0
\end{align}
\end{subequations}
We proceed with the second row of $D$. Since $D$ is symmetric, we calculate only terms of the upper triangular matrix. From \eqref{g0c}:
\begin{subequations}
\label{djr1}
\begin{align}
\label{g0c22}
D_{11}(z)&= T^\text{av}_t\left\{g_{10}(z,t)g_{10}(z,t)\right\} + T^\text{av}_t\left\{g_{11}(z,t)g_{11}(z,t)\right\}\\
&=\frac{\sigma^2}{2A^2}\left(-s_\chi+ c_\chi \tan(2\chi) \right)^2+\frac{\sigma^2}{2A^2}\left(c_\chi+s_\chi\tan(2\chi)\right)^2\\
&=\frac{\sigma^2}{2A^2}\left[s_\chi^2+c_\chi^2 \tan^2(2\chi)-2s_\chi c_\chi \tan(2\chi) + c_\chi^2+s_\chi^2\tan^2(2\chi)+2s_\chi c_\chi \tan(2\chi)\right]\\
&=\frac{\sigma^2}{2A^2}\left(1+\tan^2(2\chi)\right)=\frac{\sigma^2}{2A^2}\frac{1}{\cos^2(2\chi)}\\
\label{g0c23}
D_{12}(z)&= T^\text{av}_t\left\{g_{10}(z,t)g_{20}(z,t)\right\} + T^\text{av}_t\left\{g_{11}(z,t)g_{21}(z,t)\right\}\\
&=\frac{\sigma^2}{2A^2}\left[(-s_\chi+c_\chi\tan(2\chi))\left(-\frac{c_\chi}{\cos(2\chi)}\right) + (c_\chi+s_\chi\tan(2\chi))\left( -\frac{s_\chi}{\cos(2\chi)} \right)\right]\\
&=\frac{\sigma^2}{2A^2}\left[-c_\chi^2\frac{\sin(2\chi)}{\cos^2(2\chi)}-s_\chi^2\frac{\sin(2\chi)}{\cos^2(2\chi)}\right]=-\frac{\sigma^2}{2A^2}\frac{\sin(2\chi)}{\cos^2(2\chi)}\\
\label{g0c24}
D_{13}(z)&= T^\text{av}_t\left\{g_{10}(z,t)g_{30}(z,t)\right\} + T^\text{av}_t\left\{g_{11}(z,t)g_{31}(z,t)\right\}=0
\end{align}
\end{subequations}
We proceed with the third row of $D$. Since $D$ is symmetric, we calculate only terms of the upper triangular matrix. From \eqref{g0c}:
\begin{subequations}
\label{djr2}
\begin{align}
\label{g0c33}
D_{22}(z)&= T^\text{av}_t\left\{g_{20}(z,t)g_{20}(z,t)\right\} + T^\text{av}_t\left\{g_{21}(z,t)g_{21}(z,t)\right\}\\
&=\frac{\sigma^2}{2A^2}\frac{c_\chi^2}{\cos^2(2\chi)}+\frac{\sigma^2}{2A^2}\frac{s_\chi^2}{\cos^2(2\chi)}=\frac{\sigma^2}{2A^2}\frac{1}{\cos^2(2\chi)}\\
\label{g0c34}
D_{23}(z)&= T^\text{av}_t\left\{g_{20}(z,t)g_{30}(z,t)\right\} + T^\text{av}_t\left\{g_{21}(z,t)g_{31}(z,t)\right\}=0
\end{align}
\end{subequations}
On the fourth row we need to calculate just the last term. From \eqref{g0c}:
\begin{subequations}
\label{djr3}
\begin{align}
\label{g0c44}
D_{33}(z)&= T^\text{av}_t\left\{g_{30}(z,t)g_{30}(z,t)\right\} + T^\text{av}_t\left\{g_{31}(z,t)g_{31}(z,t)\right\}\\
&=\frac{\sigma^2}{2A^2}s_\chi^2+\frac{\sigma^2}{2A^2}c_\chi^2=\frac{\sigma^2}{2A^2}
\end{align}
\end{subequations}
Substituting \eqref{djr0},\eqref{djr1},\eqref{djr2},\eqref{djr3} into the expression \eqref{g0c} for the matrix $D$, and exploting the symmetry of $D$, we obtain:
\begin{align}
\label{eqDD22}
D=\frac{\sigma^2}{2A^2}\left[\begin{array}{cccc}
1 & 0 & 0 & 0	\\
0 & 1/\cos^2(2\chi) & -\sin(2\chi)/\cos^2(2\chi) & 0	\\
0 & -\sin(2\chi)/\cos^2(2\chi) & 1/\cos^2(2\chi) & 0	\\
0 & 0 & 0 & 1
\end{array}\right]
\end{align}
Now that $D$ is known, we look for a solution of the nonlinear equation \eqref{g0b2}. One such solution for $B=4$ is
\begin{align}
\label{dasdasr13fe}
h=\frac{\sigma}{\sqrt{2} A}\left[\begin{array}{cccc}
1 & 0 & 0 & 0	\\
0 & 1 & -\tan(2\chi) & 0	\\
0 & 0 & 1/\cos(2\chi) & 0	\\
0 & 0 & 0 & 1
\end{array}\right]
\end{align}
The noise on the averaged equations is additive in the equations for $A$ and $\chi$ (first and last row in \eqref{dasdasr13fe}), and multiplicative in the equations for $n\theta_0$ and $\varphi$ (second and third row in \eqref{dasdasr13fe}).

\subsection{Collection of the intermediate results in the final equation} 
\label{sectog}
From \eqref{ben4290342} for $v=0$ we have
\begin{subequations}
\begin{align}
m_0 = m_{0}^{\text{from}\,f}+m_{0}^{\text{from}\,g}=T^\text{av}_t\left\{f^{(q)}_0\right\}+\frac{\sigma^2}{4A^2}
\end{align}
where in the second passage \eqref{pp1} and \eqref{eqm1g} were substituted.
From \eqref{ben4290342} for $v=1$ we have
\begin{align}
\nonumber
m_1 &= m_{1}^{\text{from}\,f}+m_{1}^{\text{from}\,g}\\
&=\left(T^\text{av}_t\left\{f^{(q)}_1\right\}+T^\text{av}_t\left\{f^{(\omega)}_1\right\}\right)-\left(T^\text{av}_t\left\{f^{(q)}_2\right\}+T^\text{av}_t\left\{f^{(\omega)}_2\right\}\right)\tan(2\chi)
\end{align}
where in the second passage \eqref{eqm2g}, \eqref{pp2} and \eqref{eqlastlast} were substituted in this order. From \eqref{ben4290342} for $v=2$ we have
\begin{align}
m_2 = m_{2}^{\text{from}\,f}+m_{2}^{\text{from}\,g}=\left(T^\text{av}_t\left\{f^{(q)}_2\right\}+T^\text{av}_t\left\{f^{(\omega)}_2\right\}\right)/\cos(2\chi)
\end{align}
where in the second passage \eqref{m3}, \eqref{pp3} and \eqref{eqlastlast} were substituted. From \eqref{ben4290342} for $v=3$ we have
\begin{align}
m_3 = m_{3}^{\text{from}\,f}+m_{3}^{\text{from}\,g}=-T^\text{av}_t\left\{f^{(q)}_3\right\}-\frac{\sigma^2}{4 A^2}\tan(2\chi)
\end{align}
\end{subequations}
where in the second passage \eqref{mqw4}, \eqref{pp4} and \eqref{eqlastlast} were substituted. The system of averaged equations \eqref{okok}, with $z_v$ defined in \eqref{zssspob} reads:
\begin{subequations}
\begin{align}
\label{eqr441}
(\ln A)'&=T^\text{av}_t\left\{f^{(q)}_0\right\}+\frac{\sigma^2}{4A^2} + \frac{\sigma}{\sqrt{2}A}\mu_0\\
\nonumber
n\theta_0'&=\left(T^\text{av}_t\left\{f^{(q)}_1\right\}+T^\text{av}_t\left\{f^{(\omega)}_1\right\}\right)-\left(T^\text{av}_t\left\{f^{(q)}_2\right\}+T^\text{av}_t\left\{f^{(\omega)}_2\right\}\right)\tan(2\chi)+\\
\label{kkio}
&\qquad\frac{\sigma}{\sqrt{2}A}\mu-\frac{\sigma}{\sqrt{2}A}\tan(2\chi)\mu_2
\\
\label{o3}
\varphi'&=\frac{T^\text{av}_t\left\{f^{(q)}_2\right\}+T^\text{av}_t\left\{f^{(\omega)}_2\right\}}{\cos(2\chi)} + \frac{\sigma}{\sqrt{2}A}\frac{\mu_2}{\cos(2\chi)}\\
\label{o4}
\chi'&=-T^\text{av}_t\left\{f^{(q)}_3\right\}-\frac{\sigma^2}{4 A^2}\tan(2\chi)-\frac{\sigma}{\sqrt{2}A}\mu_3
\end{align}
\end{subequations}
We now undo the steps of section \S\ref{sec_adapt} where we made the system to fit the structure needed for the stochastic averaging. We multiply \eqref{o3} by $\cos(2\chi)$ and \eqref{o4} by $-1$:

\begin{align}
\label{o3p}
\varphi'\cos(2\chi)&=T^\text{av}_t\left\{f^{(q)}_2\right\}+T^\text{av}_t\left\{f^{(\omega)}_2\right\}+ \frac{\sigma}{\sqrt{2}A}\mu_2\\
\label{o4p}
-\chi'&=T^\text{av}_t\left\{f^{(q)}_3\right\}+\frac{\sigma^2}{4 A^2}\tan(2\chi)+\frac{\sigma}{\sqrt{2}A}\mu_3
\end{align}
We then add \eqref{o3} multiplied by $\sin(2\chi)$ to \eqref{kkio}:
\begin{align}
\label{eqr442}
n\theta_0'+\varphi'\sin(2\chi)&=\left(T^\text{av}_t\left\{f^{(q)}_1\right\}+T^\text{av}_t\left\{f^{(\omega)}_1\right\}\right)+\frac{\sigma}{\sqrt{2}A}\mu
\end{align}
We keep as final equations \eqref{eqr441},\eqref{eqr442},\eqref{o3p} and \eqref{o4p} for the real and $i$,$j$,$k$ imaginary components:
\begin{subequations}
\label{eqavgfinal}
\begin{align}
(\ln A)'&=T^\text{av}_t\left\{f^{(q)}_0\right\}+\frac{\sigma^2}{4A^2} + \frac{\sigma}{\sqrt{2}A}\mu_0\\
n\theta_0'+\varphi'\sin(2\chi)&=\left(T^\text{av}_t\left\{f^{(q)}_1\right\}+T^\text{av}_t\left\{f^{(\omega)}_1\right\}\right)+\frac{\sigma}{\sqrt{2}A}\mu\\
\varphi'\cos(2\chi)&=T^\text{av}_t\left\{f^{(q)}_2\right\}+T^\text{av}_t\left\{f^{(\omega)}_2\right\}+ \frac{\sigma}{\sqrt{2}A}\mu_2\\
\label{eqchichichi}
-\chi'&=T^\text{av}_t\left\{f^{(q)}_3\right\}+\frac{\sigma^2}{4 A^2}\tan(2\chi)+\frac{\sigma}{\sqrt{2}A}\mu_3
\end{align} 
\end{subequations}
We can then recompose \eqref{eqavgfinal} as a single quaternion-valued stochastic differential equation, to be interpreted in the It\^{o} sense:
\begin{align} 
\nonumber
(\ln A)'+\left(n\theta_0'+\varphi'\sin(2\chi)\right)i+\varphi'\cos(2\chi)j-\chi'k = &T^\text{av}_t\left\{f^{(q)}+f^{(\omega)}\right\}\ldots\\
\label{boumboum}
\ldots &+\frac{\sigma^2}{4A^2}\left(1+\tan(2\chi)k\right)+\frac{\sigma}{\sqrt{2}A}\mu_z
\end{align}
where $\mu_z$ is a quaternion-valued white Gaussian noise, i.e.~$\mu_z(t)=\mu_0+i\mu+j\mu_2+k\mu_3$ and each $\mu_v(t)\,,v=0,1,2,3$ are real-valued white Gaussian noise processes. Finally, substituting $T^\text{av}_t\left\{f^{(q)}\right\}$ and $T^\text{av}_t\left\{f^{(\omega)}_0\right\}$ from \eqref{trg5} and \eqref{eq_sec} into \eqref{boumboum} we obtain \eqref{bb} presented in the main text.

\bibliographystyle{elsarticle-num}
\bibliography{library}

\end{document}